\documentclass[11pt,a4paper]{article}
\pdfoutput=1
\usepackage{jheppub}
\usepackage{amsmath}
\usepackage{hyperref}
\usepackage{amssymb}
\usepackage{mathtools}
\usepackage{mathrsfs}
\usepackage{amsfonts}
\usepackage{bbm}
\usepackage{enumerate}
\usepackage{array}
\usepackage{booktabs}
\usepackage{textcomp}
\usepackage{caption}
\usepackage{cite}
\usepackage{cancel}
\usepackage{units}
\usepackage{multirow}
\usepackage{pst-all}
\usepackage{longtable}
\usepackage{relsize}

\newcommand{\tr}[0]{\ensuremath{\text{tr}}}

\title{Heterotic Mini-landscape in blow-up}

\author[a,b,c]{Nana Geraldine Cabo Bizet,}
\author[a]{Hans Peter Nilles}

\affiliation[a]{Bethe Center for Theoretical Physics\\~~Physikalisches Institut der Universit\"at Bonn, 
Nussallee 12, 53115 Bonn, Germany}
\affiliation[b]{Centro de Aplicaciones Tecnol\'{o}gicas y Desarrollo Nuclear,\\ 
CEADEN Calle 30, esq.a 5ta Ave, Miramar, 6122 La Habana, Cuba}
\affiliation[c]{Theory Group, Physics Department, European Organization for Nuclear Research \\ CERN CH-1211, Gen\`eve 23, Switzerland}

\emailAdd{nana@th.physik.uni-bonn.de,nana@ceaden.edu.cu,nilles@th.physik.uni-bonn.de}

\begin{document}

\hspace{12cm} CERN-PH-TH/2013-018
\vspace{1.5cm}

\vspace{-3cm}
\abstract{Localization properties of fields in compact extra dimensions
are crucial ingredients for string model building, particularly
in the framework of orbifold compactifications. Realistic models
often require a slight deviation from the orbifold point, 
that can be analyzed using field theoretic methods considering
(singlet) fields with nontrivial vacuum expectation values.
Some of these fields correspond to blow--up modes that represent
the resolution of orbifold singularities. Improving on previous
analyses we give here an explicit example of the blow--up of a
model from the heterotic Mini--landscape.  An exact identification of the blow--up modes at various
fixed points and fixed tori with orbifold twisted fields is given. We match the massless spectra and
identify the blow--up modes as non--universal axions of 
compactified string theory. We stress the important role of the Green--Schwarz anomaly polynomial for the description of the resolution of
orbifold singularities.
}

\arxivnumber{}
\keywords{Superstrings and Heterotic Strings, Superstring Vacua, Anomalies in Field and String Theories}
\maketitle
\flushbottom


\section{Introduction}

Heterotic orbifolds are a fertile region of the string landscape \citep{Lebedev:2006kn} in which the Minimal Supersymmetric Standard Model and Grand Unification Theories are widely encountered. They possess appealing features such as the existence of fixed sets (fixed points and fixed tori) where  twisted states are localized. Those fixed sets yield quotient space singularities. Their properties depend on the action of the local subgroup of  the orbifold group which leaves the particular set fixed. This locality can cause interesting physics \citep{Forste:2004ie,Kobayashi:2004ya,Buchmuller:2004hv,Buchmuller:2005jr} and has lead to the concept of local grand unification \citep{Nilles:2009yd,Nilles:2008gq,Buchmuller:2005sh} that has served to explore many promising models. Heterotic orbifolds give rise to discrete symmetries \citep{Kobayashi:2006wq,CKMPZS}, which explain the hierarchy between the electroweak and the unification scale \citep{Kappl:2008ie}, avoid proton decay \citep{Forste:2010pf,Lee:2010gv}, give rise to flavor symmetries \citep{Ko:2007dz} and suppress the problematic $\mu$ term \citep{Casas:1992mk,Antoniadis:1994hg,Lebedev:2007hv}.

On the other hand the biggest set of heterotic string compactifications preserving $\mathcal{N}=1$ supersymmetry in 4d are the so called Calabi--Yau (CY) manifolds. The moduli space of the metric in Calabi--Yau manifolds consists of the complex structure moduli and  the complexified K\"ahler structure moduli. There are well studied examples in which twisted states of orbifold models, which acquire vevs, smooth the singularities and can be identified with the moduli of the CY manifold \citep{Hamidi:1986vh}. This is expected because both CY and orbifold compactifications preserve $\mathcal{N}=1$ supersymmetry.  In fact, all $T^6/\mathbb{Z}_n$ orbifolds are singular limits of smooth CY manifolds \citep{Candelas:1985en,Hamidi:1986vh,Erler:1992ki,Aspinwall:1994ev,Reffert:2006du,Lust:2006zh,Nibbelink:2007rd,Nibbelink:2007pn}. In the last years there has been an intense work in understanding the transition of the heterotic string compactified on those two geometries. For the string on orbifolds the conformal field theory is exactly solvable and all interactions explicitly  computable. In contrast on a smooth CY the metric is not known and one has to rely on the topological information. In general one can not solve the conformal field theory, with the exception of certain points in the moduli space where a rational CFT description is available e.g. at the Gepner points. The way to proceed is to compactify the  effective 10d   $\mathcal{N}=1$ super Yang Mills coupled to supergravity on the CY.  In this frame the index theorems \citep{Witten:1981me,Witten:1984dg}  determine the 4d massless fermionic chiral asymmetry.

There are physical motivations to deform away from the orbifold point in moduli space. At this point there are many exotics states, additional $U(1)$ symmetries and enhanced discrete symmetries. This differs from what is found in the real world and spontaneous symmetry breaking with vevs of twisted fields can give rise to much more realistic vacua.  This breaking decouples exotics from the spectrum, reduces the abelian gauge sector and breaks partially global discrete symmetries. The partial breaking of discrete symmetries can be useful to create scale hierarchy, as the one needed for the pattern of quarks and leptons masses through a Froggatt--Nielsen mechanism \citep{Froggatt:1978nt}. In addition on the orbifold there exists an anomalous $U(1)_A$ symmetry which generates a Fayet--Iliopoulos D--term (FI), which breaks supersymmetry and can be cancelled by the vevs of twisted fields \citep{Atick:1987gy,Dine:1987xk,Font:1988mm}. The twisted fields which attain vevs can correspond to moduli of the CY geometry, which vanish at the orbifold point. At the orbifold point, the full spectrum, the interactions and the discrete symmetries can be determined. Thus, this connection can be used to extract information not known in the CY \citep{Ludeling:2012cu}.


 
The techniques of algebraic geometry in toric varieties \citep{Oda,Fulton,Hori:2003ic} have been applied to make the orbifold singularities smooth \citep{Erler:1992ki,Aspinwall:1994ev,Lust:2006zh,Nibbelink:2007rd,Nibbelink:2007pn}. This process of removing the singularity and adding exceptional divisors of finite size $\text{Vol}(E_r)$ is called  blow--up or resolution, the inverse process is called blow--down. In the work  \citep{Nibbelink:2007rd}  non--compact orbifolds singularities $\mathbb{C}^3/\mathbb{Z}_N$ as background of the heterotic superstring  were resolved. In these models an abelian gauge flux in 6d is turned on. It is parametrized by vectors with indices in the $E_8\times E_8$ Cartan subalgebra which determine the field strength of a holomorphic vector bundle. In the blow--down limit these vectors correspond to shifts on the gauge degrees of freedom of the local orbifold action.  Then, if we want to identify the heterotic orbifold as the singular limit of the CY, it is necessary to construct the vector bundle in such a way that orbifold rotations on the gauge degrees of freedom (d.o.f)  are reproduced in the blow--down limit. For the compact cases in which there are different local singularities, the blow--down of local resolutions fixes the vector bundle such that it reproduces the local shifts  \citep{Nibbelink:2008tv,Nibbelink:2009sp}.


The blow--up can be identified with the process of giving vevs to twisted fields. Using an exponential redefinition those twisted fields are interpreted  as the CY K\"ahler moduli. This observation relies on the realization of  the gauge transformation \citep{Nibbelink:2009sp} and on the fact that the K\"ahler moduli are local, measuring
the complex volume of the new cycles. Furthermore, a way of identifying those blow--up modes on the orbifold with the components of the vector bundle was proposed. This is based on the fact that the  Bianchi Identities (BI) giving a consistent gauge flux, possess strong similarities with the mass equations of the orbifold states. In the gauged linear sigma model description \citep{NibbelinkGroot:2010wm}
the mass equation appears as the anomaly cancellation condition \citep{Blaszczyk:2011ib,Blaszczyk:2011hs}. Those results, opened a way to study the transition in a more precise manner. If both, the string theory on the blow--up geometry and the orbifold  with vevs are coincident, then the massless spectrum should be  identified.


In describing the departure from the orbifold point within realistic compact orbifolds \citep{Nibbelink:2009sp,Blaszczyk:2010db,Buchmuller:2012mu} some difficulties were
encountered in the $\mathbbm{Z}_{6II}$ Mini--Landscape \citep{Lebedev:2006kn,Lebedev:2007hv,Lebedev:2008un}\footnote{ There are other realistic orbifold construction like the one presented in \citep{Kim:2007mt}.}  and the $\mathbbm{Z}_2 \times \mathbbm{Z}_2$ Blaszczyk model of \citep{Blaszczyk:2009in}.  The problems have two sources. One is the absence of a unique way to perform the toric resolution. In fact, there are many different resolutions connected by  flop transitions  \citep{Blaszczyk:2010db}. The second issue is the existence of discrete torsion \citep{Vafa:1994rv,Ploger:2007iq}, which allows for brother models and creates a further ambiguity in the identification. This occurs because the identification of the vector bundle with the local  orbifold shift is only up to lattice vectors. In this work we are able to overcome those difficulties.

 

A complementary approach to explore the transition was proposed in \citep{Nibbelink:2007ew}.
This method uses the fact that on the orbifold one encounters localized anomalies \citep{Gmeiner:2002es}  which depend on the chiral states at the fixed sets. 
On the blow--up, there exists also a localization on the cycles appearing in the resolution. Using the Green--Schwarz  anomaly polynomial \citep{Green:1984sg,Schellekens:1986xh}  one can study the transition by comparing the anomaly in the blow--up and the anomaly on the orbifold deformed by vevs. At the first sight the anomaly cancellation mechanism seems very different in both cases. On the orbifold there is only one axion needed to cancel a universal anomaly whereas in the blow--up there are many anomalous $U(1)$s and many axions which cancel them.  This can be explained by the change in the massless chiral spectrum, 
due to the field redefinitions and due to the fact that certain fields become massive in the blow--up. If the orbifold constitutes the blow--down limit of the toric CY, the anomaly polynomial encodes the complete information of that transition.


We look at the transition  from the two sides. First we match the chiral massless spectrum.  Using this identification we study the transition through the match  of the anomaly cancellation in both regions of the moduli space. In  a previous work, we studied the resolution of an MSSM like  $T^6/\mathbb{Z}_7$ orbifold model \citep{Blaszczyk:2011ig}.  This was simpler because all the exceptional divisors performing the resolutions are local, thus there is a local index theorem which allows to identify the spectrum. In addition there are no orbifold brother models, and there is a unique resolution for the local singularities (giving a unique resolution for the compact space). Nevertheless insights gained in that study apply in a modified way to our new situation.


Let us sketch now how the paper is structured. In section 2 we review the heterotic string on orbifolds and we present the  
$T^6/\mathbb{Z}_{6II}$ model that we chose as the key example. Section 3 is devoted to describe the orbifold toric resolution and the dimensional reduction of the $\mathcal{N}=1$ 10d theory on it. In section 4  we select the same resolution at all local $\mathbb{C}^3/\mathbb{Z}_{6II}$ singularities. We describe the Bianchi Identities and explain  the search for blow--up modes among the orbifold twisted fields.  As brother models are present we review the Mini--landscape models \citep{Lebedev:2006kn,Lebedev:2007hv,Lebedev:2008un} to select an appropriate one.  We have searched for candidates to blow--up modes among the twisted singlets. From this search
we present one finding. In section 5 we discuss how field redefinitions are implemented
to match the massless spectrum. In section 6 we study the matching for one set of blow--up modes.  Imposing an agreement with orbifold mass terms, the allowed redefinitions are restrictive and we find one case in which the match works perfectly. In section 7, we study the anomaly
cancelation in  4d, which constitutes an independent check of the picture.  We compute the anomaly in the orbifold deformed by vevs 
and compare it with the dimensional reduction of the 10d anomaly on the resolution.  We find agreement
and local blow--up modes are identified as non--universal axions.
The universal axion on the resolution turns out to be a mixture of the single orbifold
axion and the blow--up modes. The check helps to establish the 
vacuum away from the orbifold as the CY manifold obtained by a resolution.


\section{The Orbifold}

In this section we review orbifold compactification of heterotic string theory.  We then  present  the geometry of $T^6/\mathbb{Z}_{6II}$ and the heterotic orbifold model.

\paragraph{Heterotic string in orbifolds}
The toroidal compactification of the 10d heterotic string leads to a four dimensional theory
with $\mathcal{N}=4$ supersymmetry. It is possible to define a theory in which a symmetry of
the toroidal lattice is modded out such that the 4d supersymmetry is reduced. This constitutes an orbifold compactification.  Let us start with the six dimensional  internal space and perform the toroidal compactification by identifying points under translations in a lattice $\Gamma_6$, to obtain $T^6=\mathbb{R}^6/\Gamma_6$. Now we take an isometry group $P$ of $\Gamma_6$, and
perform a  modding of this symmetry to get $T^6/P$, $P$ is called the point group.\footnote{When
this group is (non--)Abelian the orbifold is called (non--)Abelian.} Modular invariance of the string partition function requires that the space group $S=\Gamma_6\rtimes P$ is embedded in the gauge degrees of freedom, we call this embedding $\mathfrak{g}$. Then the orbifold  is defined by \citep{Bailin:1999nk}
\begin{equation}
\Omega=\mathbb{R}^6/(\Gamma_6\rtimes P)\times \Lambda/\mathfrak{g}.\label{orbifold}
\end{equation}
In the bosonic representation of the gauge sector of the heterotic theory, $\Lambda=\Gamma_8\times\Gamma_8$
denotes the internal 16d torus.  The heterotic worldsheet fields in the internal space are the bosonic space coordinates $X^{k}(z,\bar{z})$,
the fermionic right--moving modes $\tilde{\psi}^k(\bar{z})$ and the 16d torus left--moving coordinates $X^I(z)$. 
The mentioned fields transform under the orbifold action as $X^k \rightarrow\theta^{kn}X^n+l^k, \, k=5,...,10$,
 $\tilde{\psi}^k \rightarrow \theta^{kn}\tilde{\psi}^n$ and $X^I \rightarrow X^I+ V^I+A^I ,I=1,...,16$, determining the twisted
 string boundary conditions.
 
Resuming, the orbifold action is given by $\theta\in P,\ \ l\in \Gamma_6$ and  $V,A\in \mathfrak{g}$. The gauge embedding of the orbifold action is determined by $V$ and $A$, which represent the embedding of the spatial rotations $\theta$ and lattice translations $l$, respectively. The quantities $V$ and $A$ are refered to as  shifts and Wilson lines respectively.  Wilson lines turn out to be essential in order to break the gauge symmetry down to the Standard Model \citep{Ibanez:1986tp}. As there are six internal dimensions, vectors in the toroidal lattice $\Gamma_6$ can be expressed in terms of a basis  $e_\alpha,\, \alpha=1,...,6$, such that $l=n_{\alpha}e_{\alpha}, \ \  A^I=n_{\alpha}A_{\alpha}^I$ $n_{\alpha}\in \mathbb{Z}$ , where $A_{\alpha}$ is the Wilson line corresponding to the lattice translation $e_{\alpha}$.

The space group\ \ $S=\{(\theta,l)\}$ is defined as the subset of the orbifold (\ref{orbifold}) acting on the spatial internal dimensions $X^k$. Strings will propagate in the internal space given by $\mathbb{R}^6/S$.  Worldsheet supersymmetry is preserved, because the twist commutes with the supersymmetry generator. This is ensured by the fact that the fermionic right--moving modes share the orbifold rotation.  Furthermore, important objects are the fixed sets (fixed points and fixed tori) under the orbifold action. Those are defined by $X_{\mathbf{f}}=\theta X_{\mathbf{f}}+l$ where $X_{\mathbf{f}}$ are the 6d coordinates of the internal space. The space group element $(\theta,l)$ is called the constructing element of the fixed point (tori). 
 Fixed points occur if $\det(1-\theta)\neq 0$. If the determinant vanishes we encounter fixed tori. For orbifolds generated by $\mathbb{Z}_N$  rotations that preserve the  lattice $\Gamma_6$, take the orbifold action to be of the form
\begin{eqnarray}
\theta=\exp(2\pi \text{i}(v_1 J_{45}+v_2J_{67}+v_3J_{89})),\ \ \theta\in \mathbb{Z}_N,
\end{eqnarray}
i.e. the transformation is block-diagonal in  the internal part of the Lorentz group $SO(6)$. Here we denote 
the generators of rotations in the three distinct planes  by $J_{45}$, $J_{67}$, $J_{89}$.  We can impose that $\mathcal{N}=1$
supersymmetry survives the compactification. Then, the invariance of the susy algebra generators under the orbifold action
yields the condition $\sum_i v_i=0$.


There is a beautiful conformal field theory description of orbifolds that we will not review here in detail \citep{Dixon:1985jw,Dixon:1986jc,Dixon:1986qv}. Essentially one solves the worldsheet equations of motion with the given boundary conditions and quantizes the string to obtain the twisted and untwisted oscillators. The physical states are obtained by acting with the latter on the twisted and untwisted vacua.  Here we shortly present the ingredients required to compute the massless spectrum. Let us look at the states with boundary conditions given by the constructing element 
\begin{equation}
g=\left(\theta^k,m_a e_a\right)\in S.\label{constructing}
\end{equation}
The orbifold possesses untwisted and twisted modes which correspond to strings with boundary conditions $k=0$ and $k\neq 0$ respectively. The untwisted string states with constructing element $g=(1,0)$ can be described by  $|q\rangle_R \otimes \tilde\alpha |p\rangle_L$. In that formula  $q=(q^0,q^1,q^2,q^3)$ represents the momentum of the bosonized right--moving fermion. This is a weight of the 
$SO(8)$ Lorentz symmetry group which is manifest in the light cone gauge.   The quantity $p$ denotes the left moving momentum of the
16 gauge d.o.f. and takes values in the $\Gamma_8\times \Gamma_8$ lattice, whereas $\tilde{\alpha}$ schematically denotes  the set of left moving oscillators. 
The mass shell equations for massless states are given by
\begin{equation}
\frac{(p+V_g)^2}{2} + N - 1 + \delta c=~\frac{(q+\phi_g)^2}{2} - \frac{1}{2} + \delta c=0\,.
\label{masshell}
\end{equation}
Here we have set the right oscillator numbers and the right--moving momentum to zero, to allow for massless right--movers. The phases $\phi_g = k v$ appearing in (\ref{masshell}) are called local twists.   $V_g$ represents the embedding on the gauge d.o.f. of the local constructing element $g$ in (\ref{constructing}).  The zero point energy  is given by  $\delta c = \frac{1}{2} \sum_{i=1}^3 \omega_i (1 - \omega_i)$, with $\omega_i = (\phi_g)_i \mod 1$ such that $0 \leq \omega_i < 1$ and 
the left--moving oscillator number is denoted by $N$. For twisted strings it is convenient to define the shifted left--moving momentum of the state as $P_{sh} = p+ V_g$. The weight $P_{sh}$ determines the behavior of the twisted string under gauge transformations. An analogous definition is the shifted right--moving momentum $q_{sh}=q+v_g$. Then, twisted states with constructing element $g$ can be written as $|q_{sh}\rangle_R \otimes \tilde\alpha |P_{sh}\rangle_L$. They will transform under another space group element $h$ with a phase $[P_{sh}\cdot V_h - q_{sh}\cdot \phi_h -\frac{1}{2}(V_g\cdot V_h - \phi_g\cdot \phi_h)]$. The surviving twisted spectrum is determined by imposing a trivial action under $h\in S$ if $[g,h]=0$. The surviving gauge group upon compactification is
computed by determining the $E_8\times E_8$ roots $\alpha_i$ which fulfill $\alpha_i\cdot V=\alpha_i\cdot A_{\alpha}=0$.


\paragraph{The $T^6/\mathbb{Z}_{6II}$ model}
In the work \citep{Lebedev:2006kn} a large number of models of the $E_8\times E_8$ heterotic string compactified on $T^6/\mathbb{Z}_{6II}$
was studied. There, of the order of $100$ models with the spectrum of the MSSM  were found. 
This Mini--landscape constitutes a fertile region of the space of $\mathcal{N}=1$ heterotic compactifications. 
The method they employed was to create models with local GUT gauge group at the fixed sets. 
The corresponding local GUTs had gauge groups $E_6$ and $SO(10)$.  We focus on the models 
with $SO(10)$ local GUT. In those cases, the orbifold shift is chosen to break $E_8\times E_8$ down to $SO(10)$. 
Further breaking is performed by turning on the Wilson lines $A_3\equiv A_4$ and $A_5$.  The torus lattice is the
root lattice of $G_2\times SU(3)\times SO(4)$ and a basis for it can be found in \citep{Nibbelink:2009sp}.

\begin{figure}
\centering
\includegraphics[width=1\textwidth]{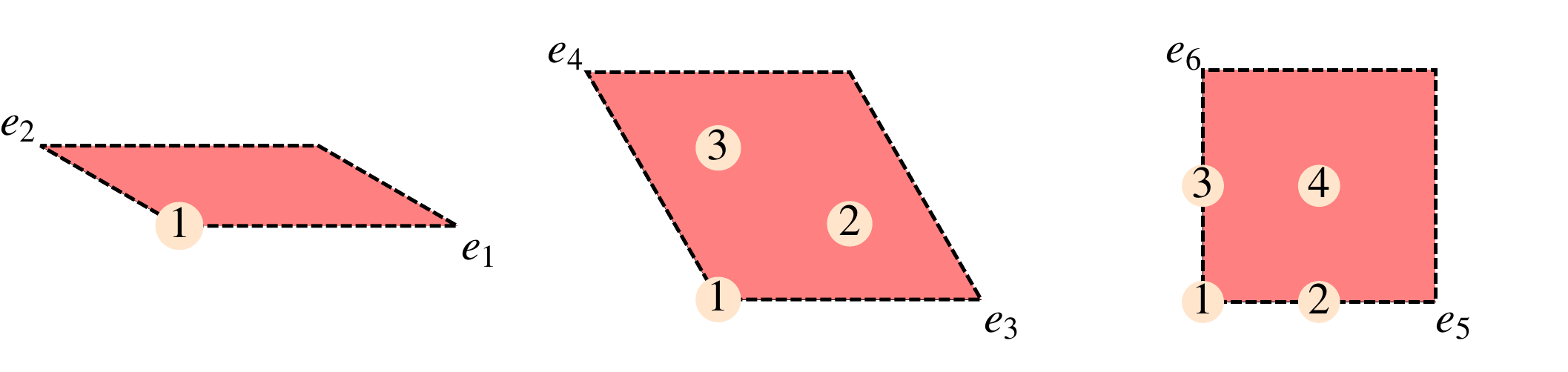}
\caption{12 fixed points of the $\theta$ sector from $T^6/\mathbb{Z}_{6II}$ orbifold. The labels of the fixed points
in the planes 1, 2 and 3 denote $\alpha,\beta$ and $\gamma$, respectively.}
\label{theta}
\end{figure}

\begin{figure}
\centering
\includegraphics[width=1\textwidth]{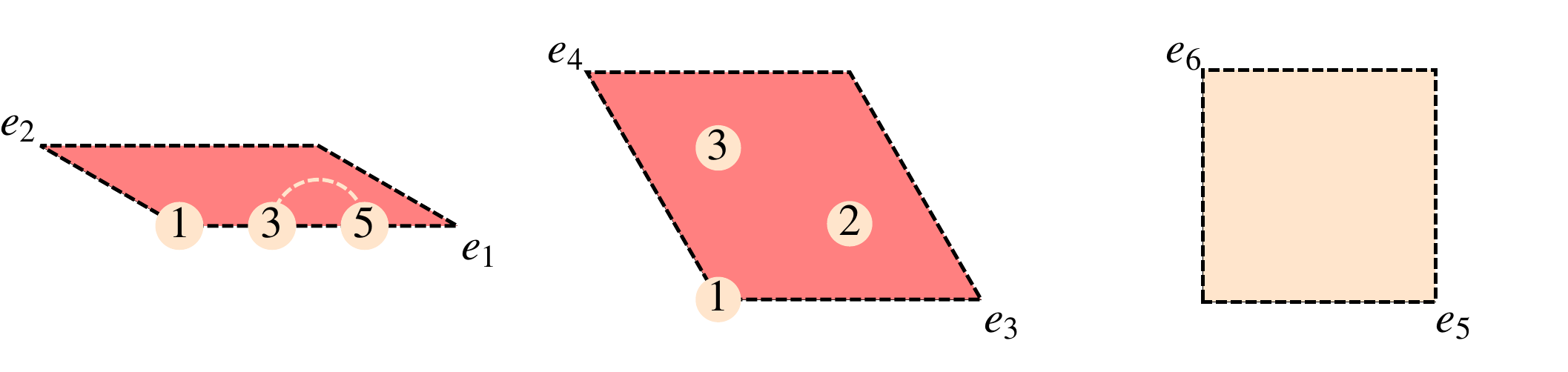}
\caption{6 fixed tori of the $\theta^2$ and $\theta^4$ sectors from $T^6/\mathbb{Z}_{6II}$ orbifold.
The labels of the fixed points in the planes 1 and 2 denote $\alpha$ and $\beta$ respectively.
Points $\alpha=3$ and $\alpha=5$ joined by a line are identified under a $\theta^3$ twist.}
\label{theta24}
\end{figure}

\begin{figure}
\centering
\includegraphics[width=1\textwidth]{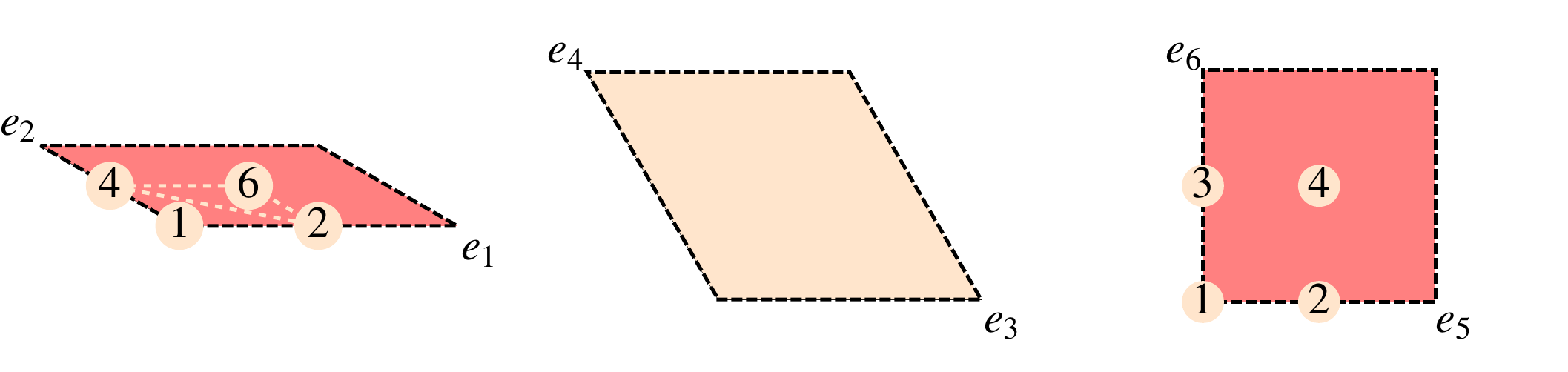}
\caption{8 fixed tori of the $\theta^3$ sector from $T^6/\mathbb{Z}_{6II}$ orbifold. The labels of the fixed points in the planes 1 and 3 denote $\alpha$ and $\gamma$ respectively. Points $\alpha=2,\, 4,\, 6$ joined by a line are identified under a $\theta^2$ twist.}
\label{theta3}
\end{figure}

In the Figures \ref{theta}, \ref{theta24} and \ref{theta3} we depict the geometry of the $T^6/\mathbb{Z}_{6II}$ orbifold.
The geometrical twist  is given by $v=\{1/6, 1/3,- 1/2\}$. Let us denote the three complex coordinates by $z_1, z_2$ and $z_3$, the twists acts on them as $\theta: z_i\rightarrow e^{2\pi i v_i}z_i$. The first figure corresponds to the first twisted sector $\theta$, which has 12 fixed points. We label the fixed points in the complex planes $i=1,2,3$ by $\alpha,\beta$ and $\gamma$ respectively, following the notation in \citep{Nibbelink:2009sp}. The Figure \ref{theta24} corresponds to the fixed tori in the $\theta^2$ and $\theta^4$ sectors. In these sectors the plane $i=3$ is a fixed torus, so the twisted states will be localized at points in the first two planes and on a torus in the third. Fixed tori with $\alpha=3,5$ are identified under the orbifold, so we have 6 fixed tori in total.  The $\theta^3$ sector is represented in Figure \ref{theta3}. In this case $z_2$ is fixed under rotations, which gives a torus in the second plane.
In the first plane the fixed tori with $\alpha=2,4,6$ are identified on the orbifold. That gives us
a total of 8 fixed tori. In Table \ref{conjclZ6II} of Appendix \ref{orbifoldtables} we give all
the conjugacy classes of this orbifold with the corresponding fixed sets, together with the labels $\alpha,\beta$ and $\gamma$ 
denoting their loci in the three complex planes. Also the Coxeter element is given.

We perform the study of the orbifold--resolution transition in Model 28 of the Mini--landscape. The shift and Wilson lines of 
that model are given by
\begin{eqnarray}
V&=&\left(\frac{1}{3},-\frac{1}{2},-\frac{1}{2},0^5,\frac{1}{2},-\frac{1}{6},-\frac{1}{2}^5,\frac{1}{2}\right),\\
A_5&=&\left( -\frac{1}{2}, -\frac{1}{2}, 0, \frac{1}{2}, \frac{1}{2}, 0, 0, 0,\frac{15}{4}, -\frac{19}{4}, -\frac{15}{4}, -\frac{15}{4}, -\frac{15}{4}, -\frac{15}{4}, -\frac{11}{4}, \frac{19}{4} \right),\nonumber\\
A_3&=&A_4=\left(\frac{1}{6},\frac{1}{6}, -\frac{1}{2}, \frac{1}{6}, \frac{1}{6}, \frac{1}{6}, \frac{1}{6}, \frac{1}{6} ,\frac{5}{3}, -\frac{2}{3}, -\frac{5}{3}, -\frac{5}{3}, -\frac{5}{3}, -\frac{5}{3}, -\frac{1}{3},\frac{8}{3}\right).\nonumber
\end{eqnarray}
The shift breaks $E_8\times E_8$ down to $SO(10)$. Adding the Wilson lines the gauge group is broken down further to $SU(3)\times SU(2)\times SU(6)\times U(1)^8$. 
A review of the non--Abelian charges of the spectrum is given in Table \ref{spectrum28}.

\begin{table}[h]
\caption{Massless spectrum of the Mini--landscape Model 28. We give the representations under the non--abelian gauge group $SU(3)\times SU(2)\times SU(6)$ and the multiplicities of the  states.}
\begin{center}
\begin{tabular}{|c|c|c|c|c|c|c|c|c|}\hline
irrep. &$\mathbf{(1,1,1)}$& $\mathbf{(1,2,1)}$& $\mathbf{(3,1,1)}$&$\mathbf{(\bar{3},1,1)}$&$\mathbf{(1,1,6)}$&$\mathbf{(1,1,\bar{6})}$&
$\mathbf{(3,2,1)}$&$\mathbf{(\bar{3},2,1)}$\\ \hline
mult.&114& 19 & 22 & 16 & 7 &  7 & 1 & 4\\ \hline
\end{tabular}
\end{center}
\label{spectrum28}
\end{table}

\section{Smoothing the singularities}
 
 In this section we describe how the spectrum is determined when compactifying the theory on the resolved space.  
 We review the resolution process of the local singularities and give the relevant data for the geometry of the resolved space. Then we describe how the dimensional  reduction of the 10d theory is performed.

\paragraph{The geometry} The local orbifold singularities are resolved and the resulting
patches are joined to obtain a global resolution \citep{Lust:2006zh}. The local singularities at the fixed points of $\theta$ are $\mathbb{C}^3/\mathbb{Z}_{6II}$ singularities. Transversally to the fixed tori of $\theta^2$ and $\theta^4$ one has $\mathbb{C}^2/\mathbb{Z}_{3}$ singularities. Similarly for the fixed tori of $\theta^3$ 
the local singularities are $\mathbb{C}^2/\mathbb{Z}_{2}$.

Let  $N_{\mathbb{R}}=\mathbb{N}^r\otimes \mathbb{R}$. A  cone $\sigma \subset N_{\mathbb{R}}$ is a set $\sigma=\{a_1v_1+a_2v_2+...+a_k v_k | a_i\in \mathbb{R},\, a_i\geq 0\}$ generated by a finite set of vectors $v_1,v_2,...,v_k$ in $\mathbb{N}^r$ such that $\sigma\cap (-\sigma)=\{0\}$. A  collection $\Sigma$ of cones in $N_{\mathbb{R}}$ is called a fan if each face of a cone in $\Sigma$ is also a cone in $\Sigma$ and the intersection of two cones in $\Sigma$ is a face of each of them. 

Starting from a fan $\Sigma$ one can construct a toric variety $X$.  The fans are spanned 
by vectors $v_1,v_2,...,v_n$ lying in the lattice $\mathbb{N}^r$. They  define a complex toric variety $X$ 
of \hbox{$\dim_{\mathbb{C}}(X)=n-r$}, as the quotient of an open subset in $\mathbb{C}^n$ under a group $G$ as $X=(\mathbb{C}^n-Z(\Sigma))/G$. 
Let us denote the coordinates by $(z_1,z_2,...,z_n)\in (\mathbb{C}^n-Z(\Sigma))$.  The vectors $v_i$ represent divisors $z_i=0$. 
The group $G$ is defined as the kernel of the map
\begin{equation}
\phi:(\mathbb{C}^*)^n\rightarrow (\mathbb{C}^*)^r,\  \ \  (t_1,....,t_n)\rightarrow (\prod_{j=1}^n t_j^{v_{j1}},...,\prod_{j=1}^n t_j^{v_{jr}}),
\end{equation}
and acts on the coordinates $z_i$ as $t_i z_i$, where $t_i$ are the solutions to $(\prod_{j=1}^n t_j^{v_{j1}},...,\prod_{j=1}^n t_j^{v_{jr}})=(1,...,1)$.  Fans describing
the d--dim local toric singularities are defined by a d-1--dim simplex $S_{d-1}$ lying in a hyperplane at distance one from the origin in $N_{\mathbb{R}}$, so that
all rays $av_k, a\in \mathbb{R}^+$ go from the origin through $S_{d-1}$. $S_{d-1}$ and its triangulation is called toric diagram in the following.  $Z(\Sigma)$ is an exclusion set encoded in the triangulation of the toric diagram \citep{Hori:2003ic} and is given by the union of divisor intersections which do not span a cone in $\Sigma$.  The variety is singular if not all the points in the lattice $\mathbb{N}^r$ can be written as a linear combination of the vectors $v_i$ with integer coefficients. Therefore, adding new vectors $\omega_r$ which subdivide the diagram is equivalent to resolving the variety.

\begin{figure}
\caption{Resolutions of the local singularities under the $\theta^3$ and $\theta^2,\theta^4$ action respectively.}
\begin{center}
\begin{tabular}{@{}cc@{}}
$\mathbb{C}^2/\mathbb{Z}_2$  &$\mathbb{C}^2/\mathbb{Z}_3$ \\
\includegraphics[height=6cm]{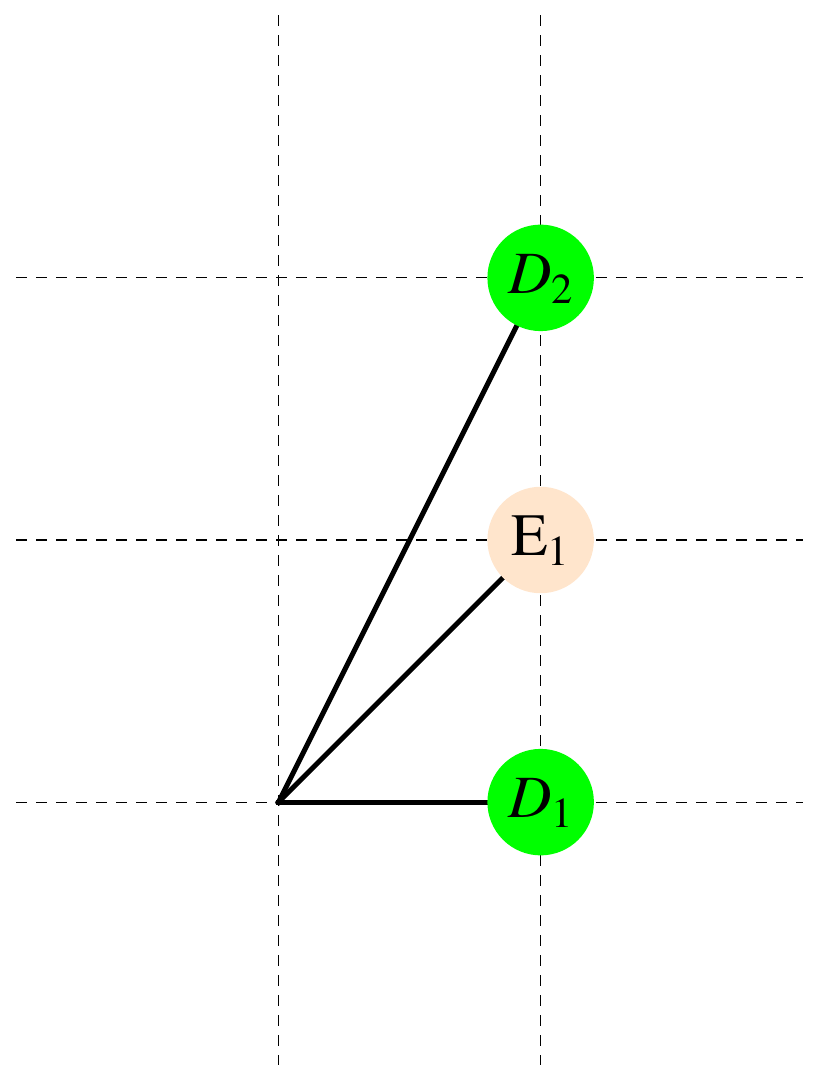} & 
\includegraphics[height=6cm]{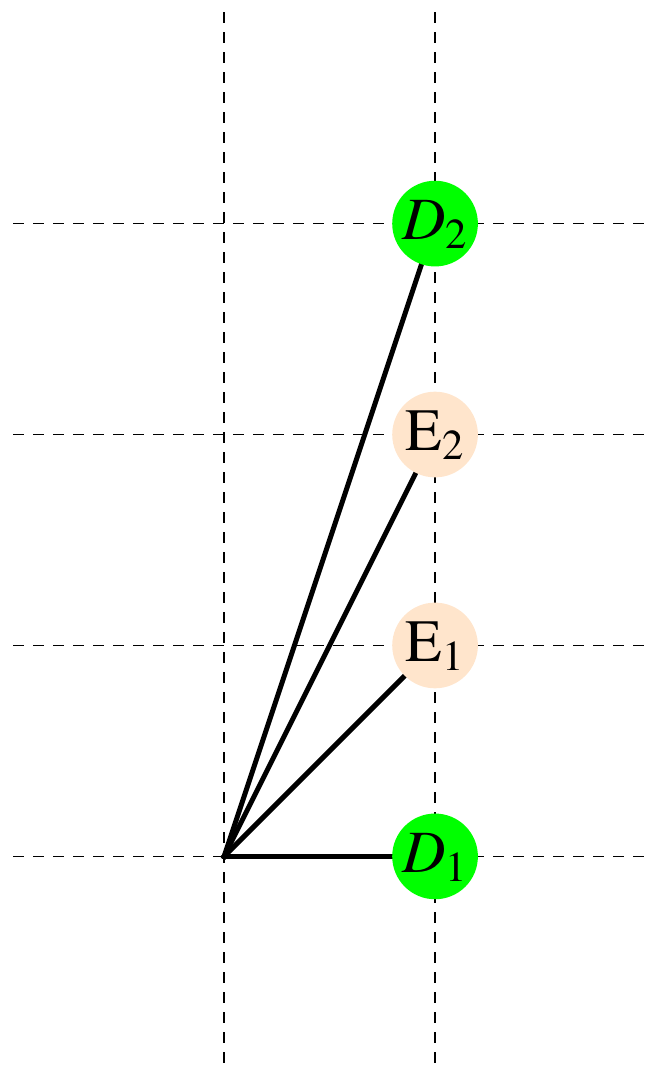} 
\end{tabular}
\end{center}
\label{toricC2Z2Z3}
\end{figure}

The orbifold singularity $\mathbb{C}^2/\mathbb{Z}_n$ has a toric diagram given by $v_1=(1,0)$ and $v_2=(1,n)$, and the
orbifold group $G=\{(t,t^{n-1}),\ \ t^n=1\}$ acts on $(z_1,z_2)$ as $(t z_1,t^{n-1} z_2)$. The divisors $D_1=\{z_1=0\}$ and $D_2=\{z_2=0\}$
correspond to the vectors $v_1$ and $v_2$. The blow--up is performed by subdividing the diagram. This is done by adding the vectors $\omega_r=(1,r),\ \  r=1,..,n-1$, which correspond to $n-1$ exceptional divisors $E_r=\{y_r=0\}$, where $y_r$ are new coordinates on the variety.  In Figure \ref{toricC2Z2Z3}, we give the
diagrams for the resolution of the local singularities $\mathbb{C}^2/\mathbb{Z}_{3}$ and $\mathbb{C}^2/\mathbb{Z}_{2}$.
The resolved $\mathbb{C}^2/\mathbb{Z}_{2}$ has coordinates $(z_1,z_2,y_1)$ identified under $G_2=\{(t,t,t^{-2}),t\in \mathbb{C}^*\}$. The exclusion set is $z_1=z_2=0$. The resolved $\mathbb{C}^2/\mathbb{Z}_{3}$ has coordinates $(z_1,z_2,y_1,y_2)$ identified under $G_3=\{(t_1,t_2,t_1^{-2}t_2,t_1t_2^{-2}),t_1,t_2\in \mathbb{C}^*\}$ with $Z(\Sigma)=(z_1=z_2=0)\cup (z_1=y_2=0)\cup (z_2=y_1=0)$.

\begin{figure}
\centering
\includegraphics[width=0.9\textwidth]{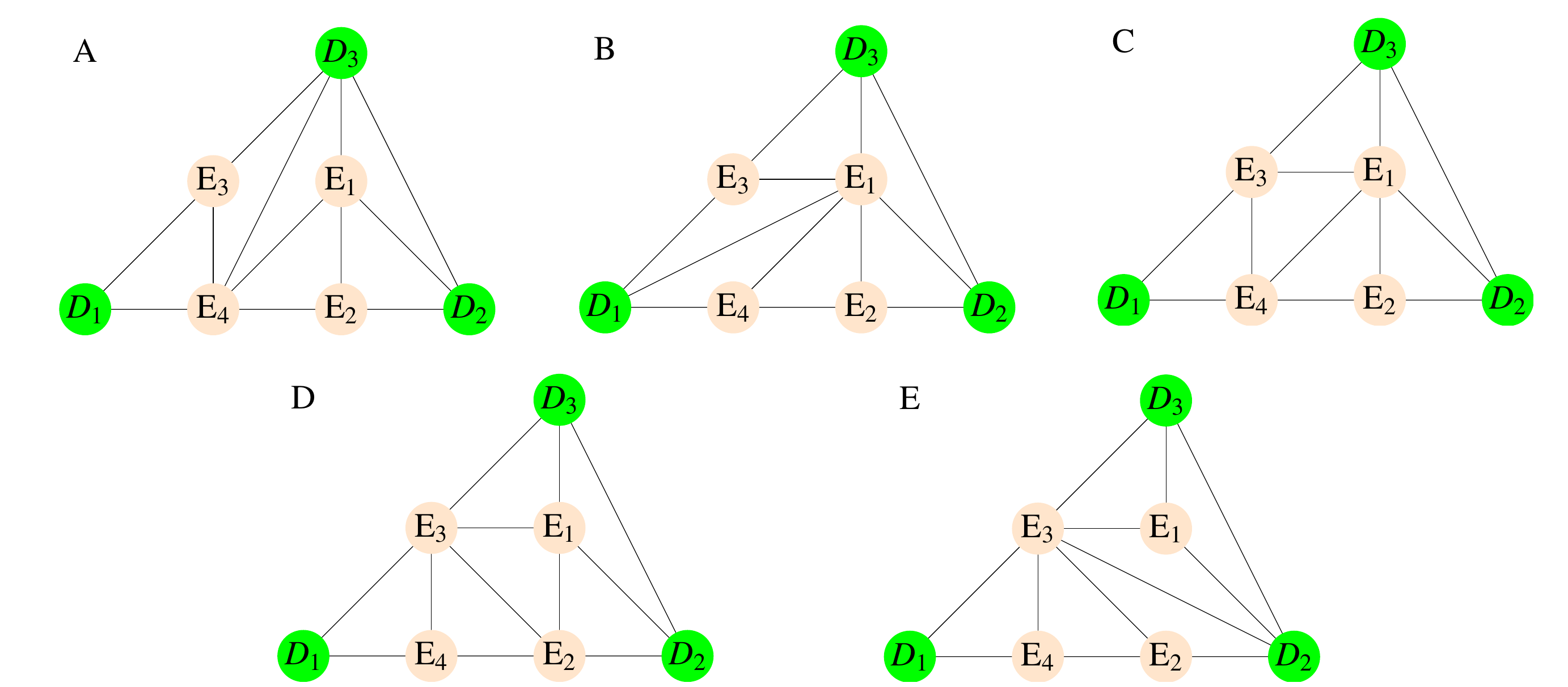}
\caption{Local resolutions of the $\mathbb{C}^3/\mathbb{Z}_{6II}$ orbifold. There are five different ways of defining
the fan \citep{Oda} which are represented by the five possible triangulations of the 2--simplices. The divisor $E_2$  corresponding to $\omega_2$ has coordinates $(0,0,1)$.
All $v_i$ are of the form $(*,*,1)$.}
\label{tria12345}
\end{figure}

Let us look at the local singularity $\mathbb{C}^3/\mathbb{Z}_{6II}$. In this case we have to add four new coordinates $y_r$ to the complex coordinates $z_1,z_2,z_3\in \mathbb{C}^3$  and four new scaling relations to define the smooth global variety.  The kernel of $(t_1,t_2,t_3,\tilde{t}_1,...,\tilde{t}_4)\rightarrow( \prod_{i,k} t^{(v_i)_1}_i\tilde{t}^{(\omega_k)_1}_k,..., \prod_{i,k} t^{(v_i)_3}_i\tilde{t}^{(\omega_k)_3}_k)$ defines the new $(\mathbb{C}^*)^4$ action. Here $\omega_k=g_iv_i$, where $z_i$ goes to $e^{2\pi i g_i v_i}z_i$ under $\theta^k$ \citep{Aspinwall:1994ev}.  Hence the new variety is defined by coordinates $(z_1,z_2,z_3,y_1,y_2,y_3,y_4)$ identified under a $(\mathbb{C}^*)^4$ action  given by
\begin{equation}
G_6=\{(t_4^{-1/6}t_5^{1/3}t_6^{-1/2}t_7^{-2/3},t_4^{-1/3}t_5^{-2/3}t_7^{-1/3},t_4^{-1/2}t_6^{-1/2},t_4,t_5,t_6,t_7), t_4, t_5,t_6,t_7\in \mathbb{C}^*\}. \nonumber
\end{equation}
The $\mathcal{N}=1$ supersymmetry condition is equivalent to the Calabi--Yau condition. The latter
is ensured if the added vectors resolving the singularity lie in the hyperplane defined by $v_1$, $v_2$ and $v_3$. The vectors $v_i$ and  $\omega_k$ are associated with divisors $D_i=\{z_i=0\}$ and $E_k=\{y_k=0\}$, which are ordinary and exceptional divisors respectively.  In the Figure  \ref{tria12345} we draw the 2--simplices $S_2$ defining the five distinct toric resolutions of $\mathbb{C}^3/\mathbb{Z}_{6II}$. 
Those are given by the different triangulations of the toric diagram. The triangulation defines the value of $Z(\Sigma)$. For example triangulation B
has $Z(\Sigma)=(y_3=y_4=0)\cup (y_3=y_2=0)$ $\cup (y_3=z_2=0)\cup (z_1=z_3=0)$ $\cup (z_1=z_2=0)\cup (z_1=y_2=0)$ $\cup (y_4=z_2=0)\cup (z_3=y_4=0)$ $\cup (z_3=y_2=0)$. Three divisors that correspond to the corners of a basic triangle have intersection 1. Triplets of divisors that do not have this property have intersection 0. Equivalence relations between the divisors are given by  $\sum(v_i)_jD_i+\sum_k (\omega_k)_jE_k\sim0$. Using Poincar\'{e} duality and Stokes theorem we relate cycles with closed--forms. Homology relations between the cycles translate into cohomology relations between the forms i.e. equivalences up to  exact cycles translates into equivalences up to exact forms. 

The global information is obtained by taking into account all local resolutions and including the inherited divisors $R_i$ which are the Poincar\'{e}  duals of the (1,1) invariant orbifold forms $dz_i\wedge d\bar{z}_{\bar{i}}$. An auxiliary polyhedron obtained in \citep{Reffert:2006du,Lust:2006zh} and employed in \citep{Nibbelink:2009sp} encodes all the triple intersections. In that way, new cohomology  classes arise in the blow--up and it is possible to determine topological information from them. Taking the volume of the resolution cycles to zero $\text{Vol}(E_r)\rightarrow 0$ the geometrical orbifold is recovered.


Applying the  method of the auxiliary polyhedra it is possible to determine the set of intersections for the compact resolved orbifold $\mathcal{M}=\widehat{T^6/\mathbb{Z}_{6II}}$. Triple intersections
of distinct divisors belonging to local resolutions have the values that can be read from the local toric diagrams. On the $\widehat{T^6/\mathbb{Z}_{6II}}$ the divisors have indices corresponding to the fixed points from which they come. $D_{i,\rho}$ represents the ordinary divisor corresponding to the fixed point singularity $\rho$ in the complex plane $i$. There are in total 10 ordinary divisors $D_{1,1}$, $D_{1,2}$, $D_{1,3}$, $D_{2,\beta}$ and $D_{3,\gamma}$. The exceptional divisors $E_{1,\beta\gamma}$, $E_{2,\alpha\beta}$, $E_{4,\alpha\beta}$ and $E_{3,\alpha\gamma}$ have their first index denoting the sector and the two following indices denoting the corresponding fixed point singularity. The global equivalence relations \citep{Nibbelink:2009sp} determine $D_{i,\rho}$ as linear combinations of $E_r$ and $R_i$. Using those equivalence relations for divisors  it is possible to obtain the non--zero intersection numbers of exceptional divisors only.  Those intersections are the ones used in performing the dimensional reduction of the 10d  theory and are given by
\begin{eqnarray}
E_{1,\beta\gamma}^3&=& 6,	E_{2,1\beta}^3 = 8,	E_{3,1\gamma}^3 = 8,	E_{4,1\beta}^3 = 8, E_{1,\beta\gamma}E_{2,1\beta}^2 = -2,\label{tria} \\
E_{1,\beta\gamma}E_{3,1\gamma}^2&=& -2, E_{1,\beta\gamma}E_{4,1\beta}^2 = -2, E_{1,\beta\gamma}E_{2,1\beta}E_{4,1\beta} = 1, E_{2,1\beta}^2E_{4,1\beta} = -2,\nonumber\\
c_2(\mathcal{M})E_{2,1\beta}&=&c_2(\mathcal{M})E_{4,1\beta}=c_2(\mathcal{M})E_{3,1\gamma}=-4,\, \, c_2(\mathcal{M})R_{2}= c_2(\mathcal{M})R_{3}=24.\nonumber
\end{eqnarray}%
The second Chern--class of the manifold $c_2(\mathcal{M})$ is the piece of degree two in the formal variables $D_J,E_r$ and $R_i$ in the total Chern--class \citep{Nibbelink:2009sp} according to
\begin{equation}
c(\mathcal{M})=\prod_{J,r}(1+D_J)(1+ E_r)(1-R_1)(1-R_2)(1-R_3)^2.
\end{equation}



\paragraph{Supergravity on the resolution}


What is known about the geometry of $\widehat{T^6/\mathbb{Z}_{6II}}$ is mainly the topological information e.g. the set of intersection numbers between divisors. In order to determine the theory in 4d one can perform a dimensional reduction of the 10d $\mathcal{N}=1$ theory, which is supergravity coupled to super Yang--Mills. Massless 4d tensor fields descend from the 10d heterotic massless tensor fields by reducing the latter on harmonic forms in $\widehat{T^6/\mathbb{Z}_{6II}}$.  Let us consider the descendants of the fields with representations $(\mathbf{35},\mathbf{1},\mathbf{1})$ and $(\mathbf{28},\mathbf{1},\mathbf{1})$ under $SO(8)\times E_8\times E_8$. Here $SO(8)$ is the Little group of the 10d Lorentz group for massless states and $E_8\times E_8$ is the gauge group of the heterotic string. The massless tensor and form fields in 10d are the metric $ \mathfrak{G}$ and the antisymmetric Kalb--Ramond field $B_2$ of $SO(8)$ respectively.  Their expansions in the base of the internal $(1,1)$ harmonic forms are \citep{Nibbelink:2009sp}\footnote{Here we use the same notation for a divisor and its dual (1,1) form.}
\begin{align}
 \mathfrak{G}=g+J = g+a_i R_i-b_{r} E_{r}\,,\qquad
 B_2 =b_2+B=b_2 + \alpha_i R_i - \beta_{r} E_{r}\,,
 \label{eq:CmpxfdKaehler}
\end{align}
where $J$ is the K\"ahler form, $B$ is the internal 6d component of $B_2$, $b_2$ is the 4d component of $B_2$ and $g$ is the 4d
component of the metric.  In four dimensions $J$ and $B$ join to form the complex scalar components of the chiral multiplets
$T_i|_{\theta=0}= a_i + i \alpha_i$ and $T_r|_{\theta=0} = b_r + i \beta_r$. The real components $a_i,b_r$  govern the size of the $R_i$ and $E_r$ cycles, respectively. The four dimensional field $b_2$ is the dual of the blow--up universal axion $a^{\rm uni}$. Let us write the field strength
$H_3$  
\begin{equation}
H_3 = d B_2 - \Omega_{3}^{\rm YM} + \Omega_{3}^{\rm L}.
\end{equation}
$\Omega_{3}^{\rm YM}$ and $\Omega_{3}^{\rm L}$ are the gauge and gravitational Chern--Simons 3--forms respectively \citep{GSW2}. The gauge invariance of $H_3$ under abelian gauge transformations with gauge parameter $\chi^ I$ implies  the following variations 
\begin{equation}
\delta \beta_{r} = V^I_{r}\chi^ I, \ \ \ \ \delta \alpha_i=0,
\end{equation}
of the dimensionally reduced antisymmetric tensor. The gauge variation of those axions cancels the 4d anomaly, which can be determined
by dimensional reduction or direct evaluation \citep{Blaszczyk:2011ig}. It is precisely the gauge variation of the $\beta_r$ moduli that leads to the interpretation that the $\exp (\beta_r)$ correspond to twisted orbifold states \citep{Nibbelink:2009sp}.


The corresponding blow--up model to a deformed orbifold has a gauge group determined by the breaking of the orbifold gauge group
by the vevs of the blow--up modes. Thus, in the blow--up the $E_8\times E_8$ group should be broken. The 6d flux $\mathcal{F}$ which
is required for consistency breaks $E_8\times E_8$ down to a subgroup.  Consistency conditions restricting the flux are the zero supersymmetric  variation of the gaugino  giving the Donaldson Uhlenbeck Yau (DUY) theorem $\int_{\mathcal{M}} \tr \mathcal{F} \wedge J \wedge J=0$ and  the Bianchi Identities $\int_S dH=0$, for
$S$ any compact divisor on the manifold $\mathcal{M}$. Abelian gauge fluxes satisfying the DUY theorem are the field strengths of holomorphic vector bundles. The  Bianchi Identities are given by
\begin{equation}
 0 = \int\limits_S\left( \tr \mathcal R^2 - \tr \mathcal F^2\right)\,, \qquad S \in \{ E_r , R_i \} \,.
 \label{BI}
\end{equation}
The second Chern--class is related to the internal curvature by $\tr \mathcal R^2=-2c_2(\widehat{T^6/\mathbb{Z}_{6II}})$. We consider an Abelian flux 
\begin{equation}
 \mathcal F = H_I V^I_r E_r, \label{bundle}
\end{equation}
where $r$ runs over indices of the orbifold fixed sets, which are denoted by $(1,\beta,\gamma)$, $(2,\alpha,\beta)$, $(4,\alpha,\beta)$ and $(3,\alpha,\gamma)$. The first entry represents the twisted sector and the  $\alpha,\beta$ and $\gamma$ indices the fixed point in the first, second and third plane. The $H_I$ are the Cartan generators of $E_8\times E_8$. The vectors $V^I_r$ determine the field strength of the vector bundle and are subject to the following constraints: they must satisfy  flux quantization conditions  which are fulfilled by requiring $V_r\sim V_{(\theta^k,\lambda)}$, where $V_{(\theta^k,\lambda)}$ is the local orbifold shift corresponding to the constructing element $(\theta^k,\lambda)$ which coincides with $r$.  Note that the above equivalence is up to lattice vectors.  The $V_r$ also have to satisfy the Bianchi Identities (\ref{BI}), which constrain their lengths and scalar products.

The massless fermion fields whose reductions gives the 4d massless chiral matter are the 10d states  $(\mathbf{8},\mathbf{248},\mathbf{1})$
and $(\mathbf{8},\mathbf{1},\mathbf{248})$ characterized by the $E_8\times E_8$ root vectors.  Their 4d multiplicity is determined using an index theorem which detects the 4d fermionic chiral asymmetry \citep{Witten:1984dg,GSW2}. The multiplicity operator is given by
\begin{align}
 \label{eq:multiplicityOperator}
\hat{N}=\frac16 \int\limits_{\mathcal{M}} \left( \mathcal{F}^3-\frac{1}{4}\tr\mathcal{R}^2 \mathcal{F}\right)\,.
\end{align}
Upon dimensional reduction (\ref{eq:multiplicityOperator}) defines with which multiplicity the states appear in the spectrum.  The surviving gauge group is determined by all the $E_8\times E_8$ roots with fulfill $\alpha_i\cdot V_r=0$. This can be seen by dimensionally reducing the Yang--Mills action.  The representation of a given state can be computed with the Dynkin labels, which are determined by the product $\alpha_i\cdot p$ between the surviving simple roots in 4d $\alpha_i$ and the root vector 
$p$ characterizing the state.

\section{Identifying the blow--up modes}

In this section we explain the search for blow--up modes using the Bianchi Identities on the resolved orbifold. We focus on the case in which all local fixed sets have the same resolution. The search for blow--up modes is performed among all twisted states of the Mini--landscape MSSM model.  We start with an orbifold model in which localized chiral superfields appear at all the fixed sets.  Then, we explore solutions of the Bianchi Identities, which correspond to massless non--oscillatory blow--up modes.  In triangulation B there are multiple sets of modes which fulfill the Bianchi Identities and therefore multiple ways of blowing--up.  They determine to which extent the hidden gauge group
is broken.  All of the encountered vacua possess moduli with different chirality in the orbifold theory.   We find
multiple solutions which correspond to massless and non oscillatory states, whose vevs preserve the hidden group.


We can fix the topology of the resolved manifold by specifying the triangulation at all local $\mathbb{C}^3/\mathbb{Z}_{6II}$ resolutions. Then, using the Bianchi 
Identities we search for consistent sets of vevs of the twisted fields.  Let us describe as an example $\widehat{T^6/\mathbb{Z}_{6II}}$ were we chose at all
fixed points triangulation B. This triangulation has the least non--vanishing self--intersections,  and therefore leads to the least restrictive equations for the vectors $V_r$.  

In $\widehat{T^6/\mathbb{Z}_{6II}}$ the abelian field strength of the vector bundle (\ref{bundle}) is given by 
\begin{equation}
\mathcal{F}=H_I\left(\sum_{\beta=1}^3\sum_{\gamma=1}^4V^I_{1,\beta\gamma} E_{1\beta\gamma}+\sum_{k=2,4}\sum_{\alpha=1,3}\sum_{\beta=1}^3V^I_{k,\alpha\beta} E_{k\alpha\beta}+\sum_{\alpha=1}^2\sum_{\gamma=1}^4V^I_{3,\alpha\gamma} E_{3\alpha\gamma}\right).
\end{equation}
To obtain the Bianchi Identities (\ref{BI}) and the multiplicity of the massless states in blow--up we need all the self intersections of exceptional divisors in equation (\ref{tria}).  The multiplicity of a 
state with $E_8\times E_8$  weight $\omega$ can be written as
\begin{eqnarray}
\hat{N}&=&\sum_{\beta\gamma}\hat{N}_{1,\beta\gamma}(\omega)+\sum_{\beta}\hat{N}_{2,\beta}(w)+\sum_{\gamma}\hat{N}_{3,\gamma}(\omega),\label{localmult}\\
\hat{N}_{1,\beta\gamma}(\omega)&=& -V_{1,\beta\gamma}\cdot\omega \left((V_{2,1\beta}\cdot w)^2 + (V_{4,1\beta}\cdot\omega)^2 - (V_{2,1\beta}\cdot w) 
(V_{4,1\beta}.\omega) - (V_{1,\beta\gamma}\cdot \omega)^2 + (V_{3,1\gamma}\cdot\omega)^2\right),\nonumber \\
\hat{N}_{2,\beta}(w)&=& \frac{1}{3} \left(4 (V_{2, 1\beta}\cdot\omega)^3 + 4 (V_{4,1\beta}\cdot\omega)^3 - 
     V_{2, 1 \beta}\cdot\omega - V_{4, 1\beta}\cdot\omega - 3 (V_{2, 1\beta}\cdot\omega)^2 (V_{4,1\beta}\cdot\omega)\right),\nonumber\\
\hat{N}_{3,\gamma}(\omega)&=& \frac{1}{3} \left(4 (V_{3,1\gamma}\cdot\omega)^3 - V_{3,1\gamma}\cdot\omega\right).\nonumber
\end{eqnarray}
Using (\ref{tria}) the Bianchi Identities (\ref{BI}) give rise to the formulas
\begin{eqnarray}
&&24-\sum_{\gamma}V^2_{3,1\gamma}-3\sum_{\gamma}V^2_{3,2\gamma}=0 \label{BI11},\\
&&3 V^2_{1,\beta\gamma}-(V_{2,1\beta};V_{4,1\beta})-V^2_{3,1\gamma}=0\label{BI12},\\
&&-2-V_{3,1\gamma}\cdot \sum_{\beta}V_{1,\beta\gamma}+2 V^2_{3,1\gamma}=0\label{BI13},\\
&&24-\sum_{\beta}(V_{2,1\beta};V_{4,1\beta})-2\sum_{\beta} (V_{2,3\beta};V_{4,3\beta})=0\label{BI14},\\
&&-12-3 V_{4,1\beta}\cdot \sum_{\gamma}V_{1,\beta\gamma}+6V^2_{4,1\beta}+2(V_{2,1\beta};V_{4,1\beta})=0\label{BI15},\\
&&-12-3 V_{2,1\beta}\cdot \sum_{\gamma}V_{1,\beta\gamma}+3V^2_{4,1\beta}+4(V_{2,1\beta};V_{4,1\beta})=0,\label{BI16}
\end{eqnarray}
for triangulation B in all local $\mathbb{C}^3/\mathbb{Z}_{6II}$ resolutions. We use the notation $ (V_1;V_2) = V_1^2+V_2^2-V_1\cdot V_2$. This set of equations allows us to explore if a given orbifold model has candidates for blow--up modes fulfilling the Bianchi Identities, which
are conjectured to be linked to the orbifold mass equations \citep{Nibbelink:2009sp,Blaszczyk:2011ib,Blaszczyk:2011hs}. This exploration can be performed in a reasonable computing time. We take the orbifold {\it Model 28} of the Mini--landscape. The equation (\ref{BI11}) is  automatically satisfied for all the corresponding states in the considered model and they are also satisfied for the Mini--landscape model discussed in \citep{Nibbelink:2009sp}.

The equations (\ref{BI11}) involving $V_{3,\alpha\gamma}$ are automatically satisfied because in all the fixed tori $(3,\alpha,\gamma)$ the singlets surviving the orbifold projection fulfill $P_{sh}^2=V^2_{3,\alpha\gamma}=\frac{3}{2}$, i.e. have zero oscillator number. 

Let us describe here the further steps of the exploration for triangulation B. First, for given values of $V_{3,\alpha\gamma}$, we select 
all the $V_{1,\beta\gamma}$ which obey (\ref{BI13}). For a fixed $V_{3,\alpha\gamma}$ there are $2401$ possibilities for  $V_{1,\beta\gamma}$. There are $50400$ $V_{2,\alpha\beta}$ and $V_{4,\alpha\beta}$ that satisfy (\ref{BI14}). From this surviving set we explore which  $V_{1,\beta\gamma},V_{2,\alpha\beta},V_{4,\alpha\beta}$ satisfy the equations (\ref{BI12}) (\ref{BI15}) and (\ref{BI16}), which turn to be the hardest to obey. An exploration for a fixed $V_{3,\alpha\gamma}$ requires $1.2\times 10^8$ iterations, while a full exploration will require of the order of $3\times 10^{10}$ iterations.  In the exploration we performed, we found multiple sets of blow--up modes which can be identified with twisted states of Model 28.

The blow--up modes identified with twisted fields  have to acquire vevs to ensure {\it D-flatness} and {\it F-flatness} of the superpotential.  The local shift  of the blow--up modes $P_{sh}\equiv V_r$ satisfy the Bianchi Identities of the vector bundle \citep{Nibbelink:2009sp}. In the following we give the simplified Bianchi Identities for triangulation B and  one of the encountered solutions.


\paragraph{Abelian vector bundles for triangulation B}

Now we come to the solutions of the Bianchi Identities for triangulation B. Considering massless and non--oscillatory modes, the equations (\ref{BI11})-(\ref{BI16}) are given by

\begin{eqnarray}
(V_{2,1\beta};V_{4;1\beta})=\frac{8}{3},\label{BI21}\\
\sum_{\beta}(V_{2,3\beta};V_{4,3\beta})=8, \label{V243b}\\
V_{3,1\gamma}\cdot \sum_{\beta}V_{1,\beta\gamma}=1,\\
V_{4,1\beta}\cdot \sum_{\gamma}V_{1,\beta\gamma}=\frac{8}{9},\\
V_{2,1\beta}\cdot \sum_{\gamma}V_{1,\beta\gamma}=\frac{10}{9}.
\end{eqnarray}
We found sets of blow--up modes which can either break or preserve the hidden group. We also explored  the possibility to obtain certain chirality features for a set of blow--up modes. The fact that the modes in sectors  $\theta^2$ and $\theta^4$ are CPT conjugate to each other is incompatible with having modes that are massless and non--oscillatory in both the $\theta$ and $\theta^3$ sectors. A  possibility would be that in the $\theta^2$ sector all the modes are left handed and in the $\theta^4$ sector all are right handed, but this can not be achieved in our case. For example in the case of $V_{2,11}$   and $V_{4,11}$ the only opposite
chirality modes are $V_{2,11}=-V_{4,11}$ and this implies $(V_{2,11};V_{2,11})=\frac{14}{3}$ which violates (\ref{BI21}). As $(2,1,1)$ is the class  conjugated  to $(4,1,1)$, this means that it is not possible to take a set of blow--up  modes in which every component of a CPT pair is identified with one blow--up mode. Having a solution in which all the blow--up modes are right or left handed is also not possible for this orbifold. For example: this restriction is seen by the fact that the fixed tori $(2,1,2)$ and $(4,1,2)$ don't possess right handed 
and left handed singlets respectively. 


The modes $V_{2,3\beta}$ and $V_{4,3\beta}$ are easily adjusted, and one can find many different solutions.
There are $107520$ solutions of equation (\ref{V243b}). If one requires that all the modes are left or right handed, there are 48 solutions. If instead one imposes that all the modes at same fixed tori from $\theta^2$ and $\theta^4$ have opposite chirality one obtains also $48$ solutions.  We focus in the set of blow--up modes given in Table \ref{blow-upT12}.  We use $F$ and $bF$ to denote left and right orbifold chiral superfields respectively and use the same notation
to denote the fermionic components of the chiral superfield, whereas we use $\langle F\rangle$ or  $\langle bF\rangle$ to denote the vevs of the 
scalar components. The non--abelian representations of the blow--up massless spectrum can be seen in Table \ref{spectrumBU}.

Similarly the solution of the BI given in \citep{Nibbelink:2009sp} has modes with different chirality. This can be checked in one of the appendices of \citep{nanathesis}, in which 
also the chirality of the twisted states are indicated. Another feature that appears in our solutions  to the BI is that  the blow--up modes can have states of equal or opposite charges in the spectrum.


\begin{table}[htdp]
\caption{Blow-up modes for triangulation B in all local $\mathbb{C}^3/\mathbb{Z}_{6II}$ resolutions. $Q_Y$ denotes
the hypercharge of the given blow--up mode.}
\begin{center}
\begin{tabular}{|c|c|c|c|c|c|}\hline
$V_r^2$ & F.P.&$Q_Y$& Numerical value of $V_r$ & irrep. & $\Phi^{\text{orb}}_{\gamma}$\\ \hline
$ \frac{25}{18} $&$ (1,1,1) $& $0$ &$ \left\{-\frac{1}{6},0,0,-\frac{1}{2},-\frac{1}{2},-\frac{1}{2},-\frac{1}{2},-\frac{1}{2},0,\frac{1}{3},0,0,0,0,0,0\right\}
$&$\mathbf{1}$&$ \text{bF57}$ \\$
 \frac{25}{18} $&$ (1,1,2) $&$\frac{1}{2}$ &$ \left\{-\frac{1}{6},0,\frac{1}{2},\frac{1}{2},\frac{1}{2},0,0,0,-\frac{1}{4},-\frac{5}{12},\frac{1}{4},\frac{1}{4},\frac{1}{4},\frac{1}{4},\frac{1}{4},-\frac{1}{4}\right\}
$&$\mathbf{1}$&$ \text{bF44}$ \\$
 \frac{25}{18} $&$ (1,1,3) $& $0$&$ \left\{-\frac{1}{6},0,0,-\frac{1}{2},-\frac{1}{2},-\frac{1}{2},-\frac{1}{2},-\frac{1}{2},0,\frac{1}{3},0,0,0,0,0,0\right\}
$&$\mathbf{1}$&$ \text{bF45}$ \\$
 \frac{25}{18} $&$ (1,1,4) $& $\frac{1}{2}$ &$ \left\{-\frac{1}{6},0,\frac{1}{2},\frac{1}{2},\frac{1}{2},0,0,0,-\frac{1}{4},-\frac{5}{12},\frac{1}{4},\frac{1}{4},\frac{1}{4},\frac{1}{4},\frac{1}{4},-\frac{1}{4}\right\}
$&$\mathbf{1}$&$ \text{bF41}$ \\$
 \frac{25}{18} $&$ (1,2,1) $& $0$ &$ \left\{-\frac{1}{2},-\frac{1}{3},0,\frac{1}{6},\frac{1}{6},\frac{1}{6},\frac{1}{6},\frac{1}{6},\frac{1}{6},\frac{1}{6},-\frac{1}{6},-\frac{1}{6},-\frac{1}{6},-\frac{1}{6},-\frac{5}{6},\frac{1}{6}\right\}
$&$\mathbf{1}$&$ \text{bF88}$ \\$
 \frac{25}{18} $&$ (1,2,2) $&$-\frac{1}{2}$ & $ \left\{0,\frac{1}{6},0,-\frac{1}{3},-\frac{1}{3},\frac{1}{6},\frac{1}{6},\frac{1}{6},\frac{5}{12},-\frac{1}{12},-\frac{5}{12},-\frac{5}{12},-\frac{5}{12},-\frac{5}{12},-\frac{1}{12},\frac{5}{12}\right\}
$&$\mathbf{1}$&$ \text{bF77}$ \\$
 \frac{25}{18} $&$ (1,2,3) $& $0$ &$ \left\{-\frac{1}{2},-\frac{1}{3},0,\frac{1}{6},\frac{1}{6},\frac{1}{6},\frac{1}{6},\frac{1}{6},\frac{1}{6},\frac{1}{6},-\frac{1}{6},-\frac{1}{6},-\frac{1}{6},-\frac{1}{6},-\frac{5}{6},\frac{1}{6}\right\}
$&$\mathbf{1}$&$ \text{bF85}$ \\$
 \frac{25}{18} $&$ (1,2,4) $&$-\frac{1}{2}$ &$ \left\{0,\frac{1}{6},0,-\frac{1}{3},-\frac{1}{3},\frac{1}{6},\frac{1}{6},\frac{1}{6},\frac{5}{12},-\frac{1}{12},-\frac{5}{12},-\frac{5}{12},-\frac{5}{12},-\frac{5}{12},-\frac{1}{12},\frac{5}{12}\right\}
$&$\mathbf{1}$&$ \text{bF70}$ \\$
 \frac{25}{18} $&$ (1,3,1) $& $0$ &$ \left\{\frac{1}{6},-\frac{2}{3},0,-\frac{1}{6},-\frac{1}{6},-\frac{1}{6},-\frac{1}{6},-\frac{1}{6},\frac{1}{3},0,-\frac{1}{3},-\frac{1}{3},-\frac{1}{3},-\frac{1}{3},\frac{1}{3},\frac{1}{3}\right\}
$&$\mathbf{1}$&$ \text{bF34}$ \\$
 \frac{25}{18} $&$ (1,3,2) $&$-\frac{1}{2}$ &$ \left\{\frac{1}{6},-\frac{2}{3},\frac{1}{2},-\frac{1}{6},-\frac{1}{6},\frac{1}{3},\frac{1}{3},\frac{1}{3},\frac{1}{12},\frac{1}{4},-\frac{1}{12},-\frac{1}{12},-\frac{1}{12},-\frac{1}{12},-\frac{5}{12},\frac{1}{12}\right\}
$&$\mathbf{1}$&$ \text{bF22}$ \\$
 \frac{25}{18} $&$ (1,3,3) $& $0$ &$ \left\{\frac{1}{6},-\frac{2}{3},0,-\frac{1}{6},-\frac{1}{6},-\frac{1}{6},-\frac{1}{6},-\frac{1}{6},\frac{1}{3},0,-\frac{1}{3},-\frac{1}{3},-\frac{1}{3},-\frac{1}{3},\frac{1}{3},\frac{1}{3}\right\}
$&$\mathbf{1}$&$ \text{bF28}$ \\$
 \frac{25}{18} $&$ (1,3,4) $&$-\frac{1}{2}$ &$ \left\{\frac{1}{6},-\frac{2}{3},\frac{1}{2},-\frac{1}{6},-\frac{1}{6},\frac{1}{3},\frac{1}{3},\frac{1}{3},\frac{1}{12},\frac{1}{4},-\frac{1}{12},-\frac{1}{12},-\frac{1}{12},-\frac{1}{12},-\frac{5}{12},\frac{1}{12}\right\}
$&$\mathbf{1}$&$ \text{bF15}$ \\$
 \frac{14}{9} $&$ (2,1,1) $& $0$ &$ \left\{-\frac{1}{3},0,1,0,0,0,0,0,0,\frac{2}{3},0,0,0,0,0,0\right\} $&$\mathbf{1}$&$ \text{bF115}$ \\$
 \frac{14}{9} $&$ (2,1,2) $& $0$ &$ \left\{\frac{1}{2},-\frac{1}{6},\frac{1}{2},-\frac{1}{6},-\frac{1}{6},-\frac{1}{6},-\frac{1}{6},-\frac{1}{6},\frac{1}{3},\frac{1}{3},-\frac{1}{3},-\frac{1}{3},-\frac{1}{3},-\frac{1}{3},\frac{1}{3},\frac{1}{3}\right\}
$&$\mathbf{1}$&$ \text{F36}$ \\$
 \frac{14}{9} $&$ (2,1,3) $& $0$ &$ \left\{-\frac{1}{6},\frac{1}{6},\frac{1}{2},\frac{1}{6},\frac{1}{6},\frac{1}{6},\frac{1}{6},\frac{1}{6},\frac{1}{6},\frac{1}{2},-\frac{1}{6},-\frac{1}{6},-\frac{1}{6},-\frac{1}{6},-\frac{5}{6},\frac{1}{6}\right\}
$&$\mathbf{1}$&$ \text{F45}$ \\$
 \frac{14}{9} $&$ (4,1,1) $& $0$ &$ \left\{-\frac{2}{3},0,0,0,0,0,0,0,0,\frac{1}{3},0,0,0,0,1,0\right\} $&$\mathbf{1}$&$ \text{bF183}$ \\$
 \frac{14}{9} $&$ (4,1,2) $& $0$ &$ \left\{\frac{1}{2},-\frac{5}{6},-\frac{1}{2},\frac{1}{6},\frac{1}{6},\frac{1}{6},\frac{1}{6},\frac{1}{6},\frac{1}{6},\frac{1}{6},-\frac{1}{6},-\frac{1}{6},-\frac{1}{6},-\frac{1}{6},\frac{1}{6},\frac{1}{6}\right\}
$&$\mathbf{1}$&$ \text{bF187}$ \\$
 \frac{14}{9} $&$ (4,1,3) $& $0$ &$ \left\{-\frac{1}{3},-\frac{2}{3},0,\frac{1}{3},\frac{1}{3},\frac{1}{3},\frac{1}{3},\frac{1}{3},-\frac{1}{6},\frac{1}{2},\frac{1}{6},\frac{1}{6},\frac{1}{6},\frac{1}{6},-\frac{1}{6},-\frac{1}{6}\right\}
$&$\mathbf{1}$&$ \text{F106}$ \\$
 \frac{14}{9} $&$ (2,3,1) $& $0$ &$ \left\{-\frac{1}{3},0,-1,0,0,0,0,0,0,\frac{2}{3},0,0,0,0,0,0\right\} $&$\mathbf{1}$&$ \text{bF97}$ \\$
 \frac{14}{9} $&$ (2,3,2) $& $0$ &$ \left\{-\frac{1}{2},\frac{5}{6},\frac{1}{2},-\frac{1}{6},-\frac{1}{6},-\frac{1}{6},-\frac{1}{6},-\frac{1}{6},-\frac{1}{6},-\frac{1}{6},\frac{1}{6},\frac{1}{6},\frac{1}{6},\frac{1}{6},-\frac{1}{6},-\frac{1}{6}\right\}
$&$\mathbf{1}$&$ \text{bF90}$ \\$
 \frac{14}{9} $&$ (2,3,3) $& $0$ &$ \left\{-\frac{2}{3},-\frac{1}{3},0,-\frac{1}{3},-\frac{1}{3},-\frac{1}{3},-\frac{1}{3},-\frac{1}{3},\frac{1}{6},-\frac{1}{2},-\frac{1}{6},-\frac{1}{6},-\frac{1}{6},-\frac{1}{6},\frac{1}{6},\frac{1}{6}\right\}
$&$\mathbf{1}$&$ \text{bF103}$ \\$
 \frac{14}{9} $&$ (4,3,1) $& $0$ &$ \left\{-\frac{2}{3},0,0,0,0,0,0,0,0,\frac{1}{3},0,0,0,0,1,0\right\} $&$\mathbf{1}$&$ \text{bF165}$ \\$
 \frac{14}{9} $&$ (4,3,2) $& $0$ &$ \left\{-\frac{1}{2},\frac{1}{6},-\frac{1}{2},\frac{1}{6},\frac{1}{6},\frac{1}{6},\frac{1}{6},\frac{1}{6},-\frac{1}{3},-\frac{1}{3},\frac{1}{3},\frac{1}{3},\frac{1}{3},\frac{1}{3},-\frac{1}{3},-\frac{1}{3}\right\}
$&$\mathbf{1}$&$ \text{bF170}$ \\$
 \frac{14}{9} $&$ (4,3,3) $& $0$ &$ \left\{\frac{1}{6},-\frac{1}{6},-\frac{1}{2},-\frac{1}{6},-\frac{1}{6},-\frac{1}{6},-\frac{1}{6},-\frac{1}{6},-\frac{1}{6},-\frac{1}{2},\frac{1}{6},\frac{1}{6},\frac{1}{6},\frac{1}{6},\frac{5}{6},-\frac{1}{6}\right\}
$&$\mathbf{1}$&$ \text{bF159}$ \\$
 \frac{3}{2} $&$ (3,1,1) $& $0$ &$ \left\{0,-\frac{1}{2},\frac{1}{2},0,0,0,0,0,0,1,0,0,0,0,0,0\right\} $&$\mathbf{1}$&$ \text{bF155}$ \\$
 \frac{3}{2} $&$ (3,1,2) $&$\frac{1}{2}$ &$ \left\{\frac{1}{2},0,\frac{1}{2},\frac{1}{2},\frac{1}{2},0,0,0,\frac{1}{4},-\frac{1}{4},-\frac{1}{4},-\frac{1}{4},-\frac{1}{4},-\frac{1}{4},-\frac{1}{4},\frac{1}{4}\right\}
$&$\mathbf{1}$&$ \text{bF153}$ \\$
 \frac{3}{2} $&$ (3,1,3) $& $0$ &$ \left\{0,-\frac{1}{2},\frac{1}{2},0,0,0,0,0,0,1,0,0,0,0,0,0\right\} $&$\mathbf{1}$&$ \text{bF154}$ \\$
 \frac{3}{2} $&$ (3,1,4) $&$\frac{1}{2}$ &$ \left\{\frac{1}{2},0,\frac{1}{2},\frac{1}{2},\frac{1}{2},0,0,0,\frac{1}{4},-\frac{1}{4},-\frac{1}{4},-\frac{1}{4},-\frac{1}{4},-\frac{1}{4},-\frac{1}{4},\frac{1}{4}\right\}
$&$\mathbf{1}$&$ \text{bF150}$ \\$
 \frac{3}{2} $&$ (3,2,1) $& $0$ &$ \left\{0,\frac{1}{2},-\frac{1}{2},0,0,0,0,0,0,-1,0,0,0,0,0,0\right\} $&$\mathbf{1}$&$ \text{bF147}$ \\$
 \frac{3}{2} $&$ (3,2,2) $&$\frac{1}{2}$ &$ \left\{\frac{1}{2},0,\frac{1}{2},\frac{1}{2},\frac{1}{2},0,0,0,\frac{1}{4},-\frac{1}{4},-\frac{1}{4},-\frac{1}{4},-\frac{1}{4},-\frac{1}{4},-\frac{1}{4},\frac{1}{4}\right\}
$&$\mathbf{1}$&$ \text{bF134}$ \\$
 \frac{3}{2} $&$ (3,2,3) $& $0$ &$ \left\{0,\frac{1}{2},-\frac{1}{2},0,0,0,0,0,0,-1,0,0,0,0,0,0\right\} $&$\mathbf{1}$&$ \text{bF141}$ \\$
 \frac{3}{2} $&$ (3,2,4) $&$\frac{1}{2}$ &$ \left\{\frac{1}{2},0,\frac{1}{2},\frac{1}{2},\frac{1}{2},0,0,0,\frac{1}{4},-\frac{1}{4},-\frac{1}{4},-\frac{1}{4},-\frac{1}{4},-\frac{1}{4},-\frac{1}{4},\frac{1}{4}\right\}
$&$\mathbf{1}$&$ \text{bF126}$\\ \hline
\end{tabular}
\end{center}
\label{blow-upT12}
\end{table}%

\begin{table}[h]
\caption{Massless spectrum on the orbifold resolution. We give the representations under the non--abelian 4d gauge group $SU(3)\times SU(2)\times SU(6)$ and the multiplicities of the  states.}
\begin{center}
\begin{tabular}{|c|c|c|c|c|c|c|c|c|}\hline
irrep& $\mathbf{(1,1,1)}$& $\mathbf{(1,2,1)}$& $\mathbf{(3,1,1)}$&$\mathbf{(\bar{3},1,1)}$&$\mathbf{(1,1,6)}$&$\mathbf{(1,1,\bar{6})}$&
$\mathbf{(3,2,1)}$&$\mathbf{(\bar{3},2,1)}$\\ \hline
mult.&40 & 9 & 8& 2 & 4 &  4 & 0 & 3\\ \hline
\end{tabular}
\end{center}
\label{spectrumBU}
\end{table}


Another way to explore the orbifold--smooth CY transition is to start with a given orbifold vev configuration and ask if there exists a resolution topology which allows us to interpret the fields taking vevs as blow--up modes. To follow this strategy we created a code that finds the self-intersections for all $\sim 5^{12}$ triangulations\footnote{The exact number of inequivalent triangulations is given in \citep{Nibbelink:2009sp}.} . Then, for a given set of vevs for the twisted orbifold states, we can check whether the set of weights $P_{sh}$ can be a solution of the BI (\ref{BI}) in a given triangulation. This exploration requires too much computing time. We therefore concentrate on the  triangulation B, which gives a less restrictive set of equations, and search for compatible blow--up modes on the orbifold spectrum.

Looking at the Mini--landscape orbifold models we can ask which conditions they should obey such that they can be blown--up completely. 
The first requirement is that they have twisted matter in every fixed point or fixed torus. From the Mini--landscape models with $SO(10)$ shift and two Wilson lines this criterium is only fulfilled by 2 out of 80 models.  The fixed tori with constructing elements $(0,0,0,0,0,0), (0,0,0,0,0,1)$ in the $\theta^3$ sector are usually empty.  We understand that by looking at the orbifold projection conditions \citep{nanathesis}. The fixed tori share projection conditions with $V_h=A_3(m_3 + m_4) + kV, k = 0,...,5$. Those conditions are more restrictive than the ones of other fixed tori. For example the $\theta^3$ fixed tori $(1,0,0,0,1,0)$, $(1,0,0,0,1,1)$ involve projections under $V_h=A_3(m_3+m_4),3V+A_3(m_3+m_4)-A_5$. 

Let us comment on how the blow--up breaks the hypercharge of the model. There is a simple argument
that shows that Mini--landscape models with $SO(10)$ shift can not be blown--up completely. In those models the SM gauge group is embedded in $E_8\times E_8$ as
\begin{eqnarray}
\alpha_1&=&(0,0,0,0,0,1,-1,0),\,  \, \alpha_2=(0,0,0,0,0,0,1,-1),\,  \, \alpha_3=(0,0,0,1,-1,0,0,0),\, \label{mini} \\
Y&=&\left(0,0,0,\frac{1}{2},\frac{1}{2},-\frac{1}{3},-\frac{1}{3},-\frac{1}{3}\right).
\nonumber
\end{eqnarray}
For the Model $28$ the following equations hold
\begin{eqnarray}
(V.Y,A_5\cdot Y,A_3.Y)=(0,\frac{1}{2},0),\\
(V.\alpha_{1,2,3},A_5.\alpha_{1,2,3},A_3.\alpha_{1,2,3})=(0,0,0).\nonumber
\end{eqnarray}
Then, assume that in the fixed set with $n_5=0$ there is a blow--up mode which is neutral under the SM gauge group and has in particular zero hypercharge. In this case the left--moving momentum of the state is $P_{sh}=p_1+k V+(n_3+n_4)A_3$ implying  $p_1\cdot Y=p_1\cdot \alpha_{1,2,3}=0$. Then, let us explore if in the fixed set with conjugacy class differing just by $n_5=1$ a singlet with zero hypercharge can exist. Denote
the left--momentum by  $P_{sh,2}=p_2+k V+(n_3+n_4)A_3+A_5$ which implies  $p_2\cdot Y=-1/2,\, p_2\cdot \alpha_{1,2,3}=0$. 
Then, the quantity $\alpha_0=p_1-p_2$ has to fullfill $\alpha_0\cdot \alpha_{1,2,3}=0,\, \alpha_0\cdot Y=-1/2$.
Taking into account (\ref{mini}) we obtain that
\begin{equation}
p_1-p_2=(*,*,*,a-1/2,a-1/2,a,a,a)\notin \Gamma_8\times \Gamma_8.
\end{equation}
This contradiction means that if there exists an SM singlet with zero hypercharge in any fixed point
with $n_5=0$ then in any fixed point differing only by $n_5=1$ a hypercharge neutral singlet can not exist.
This argument is in perfect agreement with the set of blow--up modes given in Table \ref{blow-upT12}.



\section{Field redefinitions}
\label{sec:redef}

We want to test if the deviation from the orbifold vacuum produced by vevs corresponds to a smooth Calabi--Yau
manifold.  For this aim the next step after identifying the blow--up modes is to compare the massless spectrum.  The massless chiral spectrum remaining after assigning vevs  should coincide with the massless spectrum  in the heterotic supergravity coupled to super Yang--Mills on the resolved variety. A first observation
is that states on the orbifold $\Phi^{\text{orb}}_{\gamma}$ have weights $P_{sh}$ which are different from the weights of the supergravity states which belong to the $E_8\times E_8$ root lattice. For this reason field redefinitions must be performed \citep{Nibbelink:2007ew,Blaszczyk:2011ig}.  We perform redefinitions employing the blow--up modes $\Phi_i^{\text{bu--mode}}$. Those have to reproduce the chiral asymmetry of the supergravity on the blow--up. We require  that the sum of the left moving momenta of the states add up to a vector in the lattice. We  consider redefinitions of the kind
\begin{equation}
\Phi^{\text{bu}}_{\gamma}=\Phi^{\text{orb}}_{\gamma}\prod_i(\Phi_i^{\text{bu--mode}})^{c^{\gamma}_i},\ c^{\gamma}_i\in \mathbb{Z},\label{redef1}
\end{equation}
with integer coefficients  $c_i^{\gamma}$ such that the map is single valued, and where  $\Phi_{\gamma}^{\text{bu}}$  is a chiral state on the blow--up.  The constructing elements of $\Phi_{\gamma}^{\text{orb}}$ and $\Phi_i^{\text{bu--mode}}$ are given by $g=(\theta^k,n_\alpha e_\alpha)$ and $g_i=(\theta^{k_i},m^i_\alpha e_\alpha)$ respectively. One can consider different numbers of blow--up modes in one redefinition. We studied the cases involving 1,2, or 3 blow--up modes. Let us denote the root system of $E_8\times E_8$ by $\lambda$ and recall that we call $\Lambda$ the
root lattice. Then the left moving momentum of the blow--up state 
is $P_{\text{bu}}^{\gamma}\in \lambda$. We denote the left moving momentum of the twisted state and the blow--up mode 
$i$ by  $P_{sh}^{\gamma}$ and $P_{sh}^i$ respectively. They are given by
\begin{eqnarray}
P_{sh}^{\gamma}=p+k V+n_{\alpha}A_{\alpha},\\
P_{sh}^i=p_i+k_i V+m^i_{\alpha}A_{\alpha},
\nonumber
\end{eqnarray}
with $p,p_1\in \Lambda$.  The shift and Wilson lines  have to satisfy $6V,3 A_3,3A_4,2 A_5, 2 A_6 \in \Lambda$.  
The redefinition should add a momentum to $P_{sh}^{\gamma}$ such that the result is  a vector of $\lambda$.  
Given the redefinition  (\ref{redef1}) we obtain for the momentum of the blow--up state 
\begin{eqnarray}
P_{\text{bu}}^{\gamma}=&&p+\sum_ic_i^{\gamma} p_i+\delta,\label{redef2}\\
\delta&=&\left(k+\sum_i c_i^{\gamma} k_i\right)V+\left(n_{\alpha}+\sum_i c_i^{\gamma} m^i_{\alpha}\right)A_{\alpha}.\nonumber
\end{eqnarray}
The sum (\ref{redef2}) has to be in the lattice of $\Gamma_8\times \Gamma_8$. This restricts the redefinitions as follows
\begin{eqnarray}
(k+\sum_i c_i^{\gamma} k_i)&=&0\mod 6,\\
(n_{3}+\sum_i c_i^{\gamma} m^i_3+n_{4}+\sum_i c_i^{\gamma} m^i_4)&=&0\mod 3, \nonumber\\
n_{5,6}+\sum_i c_i^{\gamma} m^i_{5,6}&=&0\mod 2. \nonumber
\end{eqnarray}

In the study of $T^6/\mathbb{Z}_7$ \citep{Blaszczyk:2011ig} we allowed only for  redefinitions of fields at the same fixed points. Here the situation is more complicated, because there are not only fixed points, but also fixed tori. In addition the orbifold is factorizable,  implying that in some planes the localization of the states can be the same, even if they are not in the same fixed set.  Here, even if
two twisted states belong to different fixed sets, their localizations have a non trivial overlap.  For this reason we have to relax the local redefinition conditions. Let us write the blow--up modes as $\Phi_{(\theta^k,\alpha\beta\gamma)}$, where the index 
represents their constructing element. One example of allowed redefinitions with 3 blow--up modes is given by
\begin{eqnarray}
\Phi_{\gamma}^{\text{bu}}&=&\Phi^{\text{orb}}_{\gamma,111}\Phi_{(\theta,111)}^{-1}\Phi_{(\theta^2,233)}^{-1}\Phi_{(\theta^4,413)}^{-1}.
\end{eqnarray}
The labels denote the values of $\alpha,\beta$ and $\gamma$.  By checking the conjugacy classes in Table (\ref{conjclZ6II}) of Appendix \ref{orbifoldtables} one
can see that the redefinitions give a vector of $\Gamma_8\times\Gamma_8$ \footnote{We choose to parametrize the redefinitions using the vector  $(k_3, 3 k_4 - k_3, 2 k_5, 6 m)$ which reflects the fact that a valid redefinition is given by $\delta=(3k_4A_{3,4}+2k_5A_5+6mV)\in \Gamma_8\times\Gamma_8$, and this ensures that $P_{\text{bu}}\in \Gamma_8\times \Gamma_8$. 
 For one and two blow--up modes we computationally explore possible redefinitions with
$-3\leq k_6\leq 3,\ \ \ \  -3\leq k_4\leq 3,\ \ \ \ -2\leq k_5\leq 2,\ \ \ -1\leq m\leq 1$. For three blow--up modes we explore possible redefinitions with $-6\leq k_3\leq 6,\ \ \ \  -1\leq k_4\leq 1,\ \ \ \ -2\leq k_5\leq 2,\ \ \ -1\leq m\leq 1$.}. In the table in Appendix \ref{redefinitions} we have collected
a set of redefinitions which realizes the orbifold--resolution map.  The exploration indicates that the correct redefinitions involve blow--up modes from different fixed points.

\section{Match of the massless spectrum}

In this section we describe the identification of the massless spectrum of the
orbifold deformed by a vev configuration and the supergravity theory on the resolution. We
search for field redefinitions that reproduce the chiral asymmetry of blow--up fermions which
agree with the orbifold superpotential mass terms. We have explored the mass terms coming from 
Yukawa couplings involving blow--up modes. We don't consider higher order terms in the superpotential, 
because they are suppressed by $M_{s}$. In addition we do not  have access to the interactions in the smooth CY.  
The superpotential terms are computed with the 
Orbifolder program \citep{Nilles:2011aj} using the classical orbifold selection rules. The multiplicity 
(\ref{localmult}) determines the difference between the states mapped to the fields $\Phi_x$ and $\bar{\Phi}_x$  in blow--up. 
It is the diagonalization of the mass matrix that determines which is the surviving massless physical state. 
We denote the blow--up states with charges in the first $E_8$ 
by $I$, in the second $E_8$ by $II$, and by $III$ when they have zero multiplicity  \footnote{Only the states charged under the surviving 
gauge symmetries in the first $E_8$ can have zero multiplicity.}. A list with all the blow--up massless
states is given in Table \ref{bustates} of Appendix \ref{buspectrum}, there the fields are ordered as $\Phi^{I,II,III}$.

\paragraph{The $\mathbf{(3,2,1)}$ states}  Let us first describe the match of the $(\mathbf{3},\mathbf{2},\mathbf{1})$ and $(\bar{\mathbf{3}},\mathbf{2},\mathbf{1})$ 
states in Table \ref{(3,2,1)}.  According to the orbifold selection rules there are no mass terms arising from Yukawa couplings with
blow--up modes.


\begin{table}[h]
\caption{Orbifold--resolution map for the $(\mathbf{3},\mathbf{2},\mathbf{1})$ representation.}
\begin{center}
\begin{tabular}{|c|c|c|c|} \hline
Multip.& Blow--up state & Redefinition & irrep. \\ \hline
-2& $\Phi^I_{11}$& $(bF_{48},bF_{60})\rightarrow \Phi^I_{11}$& $(\mathbf{\bar{3}},\mathbf{2},\mathbf{1})$\\ \hline
-1& $\Phi^I_{20}$& $bF_{189}\rightarrow \Phi^I_{20}$&$(\mathbf{\bar{3}},\mathbf{2},\mathbf{1})$\\ \hline
0&$\Phi^{II}_{13}$& $bF_{93}\rightarrow \Phi^{III}_{13}$, $bF_{173}\rightarrow \bar{\Phi}^{III}_{13}$&$(\mathbf{3},\mathbf{2},\mathbf{1})$\\ \hline
\end{tabular}
 \end{center}
\label{(3,2,1)}
\end{table}%
The orbifold fields $bF_{93}$ and $bF_{173}$ are mapped to conjugate blow--up fields $\Phi^{III}_{13}$ and $\bar{\Phi}^{III}_{13}$ respectively, which form a massive pair. On blow--up there is a net number of 3 $(\mathbf{\bar{3}},\mathbf{2},\mathbf{1})$  and 
0 $(\mathbf{3},\mathbf{2},\mathbf{1})$ massless states, whereas on the orbifold (see Table \ref{spectrum28}) there are 4 $(\mathbf{\bar{3}},\mathbf{2},\mathbf{1})$ and 1 $(\mathbf{3},\mathbf{2},\mathbf{1})$. The field redefinitions in Appendix \ref{redefinitions} give multiplicities 
which match perfectly this difference.

\paragraph{The triplets} In \citep{nanathesis} we gave redefinitions for triplets and anti--triplets which test an ansatz for local multiplicity. Here
we focus in finding a map which agrees with orbifold superpotential mass terms.  

In this case the orbifold--resolution map is summarized in Table \ref{triplets}. In the Appendix \ref{redefinitions} we explicitly give  a set of redefinitions which realize
the presented map. Looking at the superpotential mass terms the case of the triplets is interesting because a new feature appears.  Let us analyze it in detail. We start by listing the mass terms in which triplets and blow--up modes are involved.  One of them is
\begin{equation}
bF_6(bF_{152}\langle bF_{153}\rangle+bF_{149}\langle bF_{150}\rangle+bF_{133}\langle bF_{134}\rangle+bF_{125}\langle bF_{126}\rangle),
\end{equation}
where $bF_6$ is an untwisted field and is exactly identified with $\Phi^{I}_{4}$. Further, we perform the redefinitions  $bF_{152},bF_{149},bF_{133},bF_{125}\rightarrow$ $\bar{\Phi}^{I}_{4}$, where $\Phi^{I}_{4}$ and $\bar{\Phi}^{I}_{4}$ are conjugate pairs. A perfect agreement with the orbifold mass terms is found. Next consider the masses
\begin{eqnarray}
bF_{112}(bF_{152}\langle bF_{77}\rangle+bF_{149}\langle bF_{70}\rangle+bF_{133}\langle bF_{77}\rangle+bF_{125}\langle bF_{70}\rangle),\\
bF_{92}(bF_{152}\langle bF_{77}\rangle+bF_{149}\langle bF_{70}\rangle+bF_{133}\langle bF_{77}\rangle+bF_{125}\langle bF_{70}\rangle).\nonumber
\end{eqnarray}
Performing the redefinitions $(bF_{112},bF_{92})\rightarrow$ $\Phi^{I}_{4}$, the counting gives one massless $\bar{\Phi}^{I}_{4}$ state in blow--up. Still, we perform a last identification of a pair without orbifold mass terms  $(bF_{121},bF_{129})\rightarrow \Phi^I_{4}$,  having in total one  $\Phi^I_{4}$ mode.

Also the following masses agree easily with redefinitions
\begin{eqnarray}
bF_{62}\langle bF_{157}\rangle(bF_{169}+bF_{185}),\\
bF_{50}\langle bF_{45}\rangle(bF_{169}+bF_{185}).\nonumber
\end{eqnarray}
They allow the identifications $(bF_{62},bF_{50})$$\rightarrow$ $\Phi^{III}_{16}$
and $(bF_{169},bF_{185})$$\rightarrow$ $\bar{\Phi}^{III}_{16} $\footnote{Given that the mass
matrix has maximal rank.}.

The mass terms
\begin{eqnarray}
bF_{20}(bF_{151}\langle F_{106}\rangle+bF_{132}\langle F_{106}\rangle+bF_{23}(\langle F_{45}\rangle+\langle bF_{159}\rangle)),\\
bF_{13}(bF_{148}\langle F_{106}\rangle+bF_{124}\langle F_{106}\rangle+bF_{16}(\langle F_{45}\rangle+\langle bF_{159}\rangle)),\nonumber
\end{eqnarray}
are redefined as shown in Table \ref{triplets}.  In the following a feature arises that has not been observed before.
Consider the remaining mass terms
\begin{eqnarray}
&&bF_{116}(bF_{30}\langle bF_{154}\rangle+bF_{36}\langle bF_{155}\rangle+bF_{125}\langle bF_{15}\rangle+bF_{133}\langle bF_{22}\rangle+\label{eigen}\\&&bF_{149}\langle bF_{15}\rangle+bF_{152}\langle bF_{22}\rangle),\nonumber
\\
&&bF_{105}(bF_{30}\langle bF_{154}\rangle+bF_{36}\langle bF_{155}\rangle+bF_{125}\langle bF_{15}\rangle+bF_{133}\langle bF_{22}\rangle+\label{eigen2}\\&&bF_{149}\langle bF_{15}\rangle+bF_{152}\langle bF_{22}\rangle).\nonumber
\end{eqnarray}
If we want to fit the previously performed redefinitions with these masses, we need to make use
of the fact that the mass eigenstates will be linear combinations of $bF_{116}$ and $bF_{105}$
or $bF_{30},bF_{36},bF_{125}$ and $bF_{133}$. Then we redefine the massive combinations denoted by
$(bF_{116}+bF_{105})_1$ and $(bF_{30}+bF_{36}+bF_{125}+bF_{133})_1$ to the conjugated states
$\Phi^I_4$ and $\bar{\Phi}^I_{4}$ respectively. Those redefinitions agree with the ones given to the remaining
four fields which are mapped to $\Phi^I_4$ and $\bar{\Phi}^I_{4}$ in (\ref{eigen}).  The massless eigenstates
$(bF_{116}+bF_{105})_2$ and $(bF_{30}+bF_{36}+bF_{125}+bF_{133})_2$ are redefined
to $\Phi_{25}^I$ and $\Phi^{III}_{32}$, agreeing with all previous redefinitions.  In order to simplify the notation in Table \ref{triplets} we substitute the massless combination constructed from $bF_{30},bF_{36},bF_{125}$ and $bF_{133}$ by $bF_{30}$. Analogously  $bF_{36},bF_{125}$ and $bF_{133}$ in the table represent the remaining mass eigenstates\footnote{There will be
 another two massless combinations $(bF_{30}+bF_{36}+bF_{125}+bF_{133})_{3,4}$ also redefined
 to $\bar{\Phi}^I_{4}$.}.

Let us conclude with the overall picture. In the orbifold there are $16\, (\mathbf{3},\mathbf{1},\mathbf{1})$
and $22 \, (\mathbf{\bar{3}},\mathbf{1},\mathbf{1})$, whereas in blow--up there are 2 triplets and 8 anti--triplets.
The redefinitions performed give a map in which 8 massive vector pairs are created and the chiral
asymmetry of the Calabi--Yau compactification is reproduced.


\begin{table}[h]
\caption{Triplets identification in agreement with superpotential mass terms.}
\begin{center}
\begin{tabular}{|c|c|c|c|} \hline
Mult.& State blow--up & irrep.&redef. \\ \hline
-3&$\Phi^I_4$&$\mathbf{3}$&$(bF_{121},bF_{129}) \rightarrow \Phi^I_4$,\, {$bF_{112},bF_{92}\rightarrow \Phi^I_{4}$} \\
&&& $bF_6\equiv  \Phi^I_{4}$, $bF_{152},bF_{149},bF_{133},bF_{125} \rightarrow \bar{\Phi}^I_{4}$ \\ 
&&&$(bF_{188},bF_{172})\rightarrow \Phi^I_4$,\\ 
&&&$(bF_{116}+bF_{105})_1\rightarrow \Phi^I_4$, $bF_{36}\rightarrow \bar{\Phi}^I_{4}$\\ \hline  
-2&$\Phi^I_7$&$\mathbf{3}$& $(bF_{135},  bF_{127})\rightarrow \bar{\Phi}^I_{7}$\\ \hline 
-2&$\Phi^I_8$&$\mathbf{\bar{3}}$&$bF_{20}\rightarrow \Phi^I_8$, $bF_{151},bF_{132},bF_{23}\rightarrow \bar{\Phi}^I_{8}$\\ \hline  
-2&$\Phi^I_{12}$&$\mathbf{3}$& $(bF_{58},  bF_{46})\rightarrow \Phi^I_{12}$\\ \hline 
-1&$\Phi^I_{25}$&$\mathbf{3}$&$bF_{99}\rightarrow \Phi^I_{25}$,\, $(bF_{116}+bF_{105})_2 \rightarrow \Phi^I_{25}$\\ 
&&&$bF_{184},bF_{166}\rightarrow \Phi^I_{25}$, $bF_{100},bF_{94}\rightarrow \bar{\Phi}^I_{25}$, $bF_{174}\rightarrow \bar{\Phi}^I_{25}$\\ \hline 
0&$\Phi^{III}_{16}$&$\mathbf{3}$&$(bF_{62},bF_{50})\rightarrow \Phi^{III}_{16}$, $(bF_{169},bF_{185})\rightarrow \bar{\Phi}^{III}_{16}$\\ \hline 
0&$\Phi^{III}_{32}$& $\mathbf{\bar{3}}$ & $bF_{148},bF_{124},  bF_{16}\rightarrow \bar{\Phi}^{III}_{32}$\\ 
&&& $bF_{13}$, $bF_{30}$, \,$bF_{161}\rightarrow \Phi^{III}_{32}$\\ \hline 
\end{tabular}
\end{center}
\label{triplets}
\end{table}%

\paragraph{The doublets} The mass terms arising at tree level are given by
\begin{eqnarray}
&&bF_{11} bF_{178}\langle bF_{118} \rangle + bF_{158} bF_{31}\langle  bF_{28} \rangle + bF_{178} bF_{31} \langle bF_{28} \rangle+ \\ 
&+& bF_{158} bF_{37} \langle bF_{34} \rangle+ bF_{178} bF_{37} \langle bF_{34} \rangle+ bF_{11} bF_{175} \langle bF_{90}\rangle.\nonumber
\end{eqnarray}
Those come from trilinear couplings agreeing with standard orbifold selection rules \citep{Nilles:2011aj}. A set of redefinitions consistent with the previous mass terms is given in Table (\ref{doubletsMass}). The mass terms are of the form $\Phi^{III}_{29} \bar \Phi^{III}_{29}$. The fields $bF_{89}$ and $bF_{111}$ form the massive linear combination $(bF_{89}+bF_{111})_1$ and the massless $(bF_{89}+bF_{111})_2$. We have chosen
the given map, because in the orbifold they have opposite charges to $bF_{175}$. In the orbifold there are 19 doublets and 10 of them form mass terms giving
a total of 9 in blow--up.

\begin{table}[h]
\caption{Doublets redefinition with correct orbifold mass terms.}
\begin{center}
\begin{tabular}{|c|c|c|c|}\hline
Mult.& State blow--up & redef. & irrep.\\ \hline
-2&$\Phi^I_{10}$&$(bF_{61},bF_{49})\rightarrow \Phi^I_{10}$&$\mathbf{(1,2,1)}$\\ \hline
-2&$\Phi^I_{16}$&$bF_{24},\,(bF_{89}+bF_{111})_2\rightarrow \bar{\Phi}^I_{16}$&$\mathbf{(1,2,1)}$\\ \hline
-2&$\Phi^I_{17}$&$(bF_{42},bF_{39})\rightarrow \bar{\Phi}^{I}_{17}$&$\mathbf{(1,2,1)}$\\ \hline
-1&$\Phi^I_{19}$&$bF_8 \equiv \Phi^I_{19}$&$ \mathbf{(1,2,1)}$\\ \hline
-1&$\Phi^I_{21}$& $bF_4\equiv\bar{\Phi}^I_{21}$& $\mathbf{(1,2,1)}$\\ \hline
-1&$\Phi^I_{23}$&$bF_{10}\equiv \Phi^I_{23}$&$\mathbf{(1,2,1)}$\\ \hline
0&$\Phi^{III}_{29}$&$bF_{11}\equiv \bar{\Phi}^{III}_{29}, bF_{12}\equiv \Phi^{III}_{29}$, $bF_{108}\rightarrow \bar{\Phi}^{III}_{29},bF_{175}\rightarrow \Phi^{III}_{29}$, &$\mathbf{(1,2,1)}$ \\ 
&&$(bF_{89}+bF_{111})_1,\,(bF_{31},bF_{37})\rightarrow \bar{\Phi}^{III}_{29},bF_{17},(bF_{158},bF_{178})\rightarrow \Phi^{III}_{29}$ & \\ \hline 
\end{tabular}
\end{center}
\label{doubletsMass}
\end{table}%

\paragraph{The sixplets} The matter charged under $SU(6)_{\text{hidden}}$ has representations $(\mathbf{1},\mathbf{1},\mathbf{6})$ and  $(\mathbf{1},\mathbf{1},\bar{\mathbf{6}})$. The map can be seen in Table \ref{six}.  In the orbifold there are 7 six--plets and 7 anti--six--plets. On blow--up there are 4 of both kinds. This agrees with the 3 blow--up mass terms that can be read in the table formed by six--plets and anti--six--plets pairs.


\begin{table}[htdp]
\caption{Orbifold-resolution identification for the $\mathbf{6}$ and $\mathbf{\bar6}$ representations of $SU(6)$. }
\begin{center}
\begin{tabular}{|c|c|c|c|}\hline
Mult.& State blow--up & redef. & irrep.\\ \hline
-1&$\Phi^{II}_{4}$&$bF_{182}\rightarrow \Phi^{II}_{4}$&$\mathbf{6}$\\ \hline
-1&$\Phi^{II}_{2}$&$bF_{2}\equiv\Phi^{II}_{2}$&$\mathbf{\bar{6}}$\\ \hline
-1&$\Phi^{II}_{9}$&\begin{tabular}{c}$bF_{9}\equiv \Phi^{II}_{9}$, $bF_{136},bF_{142}\rightarrow \bar{\Phi}^{II}_{9}$,\\ $bF_{137},bF_{143}\rightarrow \Phi^{II}_{9}$\end{tabular}&$\mathbf{\bar{6}}$\\ \hline
-1&$\Phi^{II}_{19}$&\begin{tabular}{c}$bF_{164} \rightarrow \Phi^{II}_{19}$, $bF_{157} \rightarrow \Phi^{II}_{19}$,\\ $bF_{102} \rightarrow \bar{\Phi}^{II}_{19}$\end{tabular}&$ \mathbf{6}$\\ \hline
-2&$\Phi^{II}_{14}$& $(bF_{106},bF_{117})\rightarrow\Phi^{II}_{14}$& $\mathbf{\bar{6}}$\\ \hline
-2&$\Phi^{II}_{13}$&$(bF_{86},bF_{83})\rightarrow \Phi^{II}_{13}$&$\mathbf{6}$\\ \hline
\end{tabular}
\end{center}
\label{six}
\end{table}%
The orbifold superpotential mass terms  are
\begin{equation}
 bF_9 bF_{136} \langle bF_{141} \rangle + bF_9 bF_{142} \langle bF_{147}\rangle.
 \label{mass6}
\end{equation}
Equation (\ref{mass6}) shows that a massive pair is formed from $bF_9,bF_{136}$ and $bF_{142}$.  Furthermore away from the orbifold point another two pairs form to give a net field $\Phi^{II}_9$.

\paragraph{The singlets} 


At the orbifold all the untwisted singlets are massless and they only take part in Yukawa couplings with doublets.  The twisted singlets instead have various mass terms coming from  Yukawa couplings to  blow--up modes. Those are
\begin{eqnarray}
&&bF_{160}\left(bF_{84}\langle bF_{45}\rangle+bF_{87}\langle bF_{57}\rangle+bF_{27}\langle bF_{28}\rangle+bF_{33}\langle bF_{34}\rangle\right),\\
&&bF_{180}(bF_{84}\langle bF_{45}\rangle+bF_{87}\langle bF_{57}\rangle+bF_{27}\langle bF_{28}\rangle+bF_{33}\langle bF_{34}\rangle),\\
&& bF_{40}(bF_{114}(\langle bF_{134}\rangle+\langle bF_{153}\rangle)+bF_{14}\langle bF_{187}\rangle),\\
&& bF_{43}(bF_{114}(\langle bF_{126}\rangle+\langle bF_{150}\rangle)+bF_{21}\langle bF_{187}\rangle),\\
&&bF_{146} bF_{35} \langle F_{106} \rangle,\\
&& bF_{35}( bF_{146}\langle F_{106} \rangle+bF_{107}\langle bF_{155}) \rangle,\\
&& bF_{29}(bF_{140}\langle F_{106} \rangle+bF_{107}\langle bF_{154} \rangle).
\end{eqnarray}
It is easy to check by looking at Table \ref{singlets1} that the identifications agree with the mass terms
of the orbifold superpotential.

There is an ingredient not shown in the map presented so far. In  the superpotential there are Yukawa couplings in which two blow--up modes
are involved.   We have checked up to trilinear order that the vevs can be assigned while ensuring F--flat vacua.
In addition, only a pair of twisted singlets written as massless in the map of the Table \ref{singlets1} becomes massive
due to those trilinear couplings. The map given above can be slightly modified to
also reproduce the CY chiral asymmetry\footnote{The fields $bF_{72},bF_{79}$ are the ones becoming massive
due to the trilinear couplings given. The change in the map is to make $bF_{95},bF_{96}\rightarrow \Phi^I_{18}$.}.
The number of singlets in the orbifold is 114 out of which 74 are redefined to conjugated states forming blow--up massive pairs, to give 
40 massless states in  blow--up.

This completes the matching of the heterotic string massless spectrum in the deformed orbifold
and in the toric CY.  At the level of the massless spectrum, the geometric resolution with abelian vector bundle
constitutes a blow--up of the MSSM Mini--landscape Model 28, in which the twisted singlets in Table \ref{blow-upT12} are identified
as the blow--up modes.

\begin{table}[h]
\caption{Singlets identification in agreement with superpotential mass terms.}
\begin{center}
\begin{tabular}{|c|c|c|c|c|} \hline
Mult.&States blow--up&redef. \\ \hline
$E^1_8$ spectrum I\\ \hline
$-4$&$\Phi^I_1$&$(bF_{63},bF_{66})\rightarrow \Phi^I_1$, \,$(bF_{51},bF_{54})\rightarrow \Phi^I_1$\\ \hline
$-4$&$\Phi^I_2$& $(bF_{64},bF_{52})\rightarrow \Phi^I_2$, \,$(bF_{81},bF_{75})\rightarrow \Phi^I_2$\\ \hline
$-2$&$\Phi^I_5$& \begin{tabular}{c}$bF_{123},bF_{35},bF_{29},bF_{65}, bF_{55}\rightarrow \Phi^I_5$,\\ \, $bF_{140},bF_{146},bF_{107}\rightarrow \bar{\Phi}^I_{5}$\end{tabular}\\ \hline
$-2$&$\Phi^I_6$&$bF_{98},bF_{114}$, $(bF_{21},bF_{14})\rightarrow \Phi^I_6$,\, $bF_{40},bF_{43}\rightarrow \bar{\Phi}^I_{6}$\\ \hline
$-2$&$\Phi^I_9$& $(bF_{82},bF_{74})\rightarrow \Phi^I_9$\\ \hline
$-2$&$\Phi^I_{13}$& $bF_{78},bF_{163}\rightarrow \bar{\Phi}^I_{13}$\\ \hline
$-2$&$\Phi^I_{14}$& $(bF_{25},bF_{18})\rightarrow \Phi^I_{14}$\\ \hline
$-2$&$\Phi^I_{15}$& $(bF_{130},bF_{122}), bF_{71}\rightarrow \Phi^I_{15}$, \,$bF_{73}\rightarrow \bar{\Phi}^I_{15}$\\ \hline
$-4$&$\Phi^I_{3}$& $bF_{1}\equiv \Phi^I_3$,\, $bF_{177}\rightarrow \Phi^I_{3}$, \,$bF_{190}\rightarrow \Phi^I_{3}$, $bF_{80}\rightarrow \Phi^I_{3}$\\ \hline
$-2$&$\Phi^I_{18}$& $(bF_{79},bF_{72})\rightarrow \bar{\Phi}^I_{18}$\\ \hline
$-1$&$\Phi^I_{24}$&\begin{tabular}{c}$bF_{87},bF_{84},bF_{171},(bF_{33},bF_{27})\rightarrow \Phi^I_{24}$,\\ 
$(bF_{26},bF_{19})\rightarrow \bar{\Phi}^I_{24}$,\, $bF_{95}, bF_{96},bF_{128}\rightarrow \Phi^I_{24}$,\\
\, \ $bF_{91},bF_{104},bF_{120},bF_{180},bF_{160}\rightarrow \bar{\Phi}^I_{24}$\end{tabular}\\ \hline
$-1$&$\Phi^I_{22}$& $bF_{7}\equiv \Phi^I_{22}$\\ \hline
 $E^2_8$ spectrum II\\ \hline
$-2$&$\Phi^{II}_{7}$&\begin{tabular}{c}$(bF_{47},bF_{59})\rightarrow \bar{\Phi}^{II}_{7}$,\, $bF_{3}\equiv \bar{\Phi}^{II}_{7}$,  $(bF_{38},bF_{32})\rightarrow \Phi^{II}_{7}$,\\ 
$bF_{113}, bF_{168},bF_{144}\rightarrow \Phi^{II}_{7}$\end{tabular}\\ \hline
$-2$&$\Phi^{II}_{3}$&$(bF_{76},bF_{69})\rightarrow \Phi^{II}_{3}$\\ \hline
$-4$&$\Phi^{II}_{20}$&$(bF_{53},bF_{67})\rightarrow\bar{\Phi}^{II}_{20}$, \, $(bF_{145},bF_{139})\rightarrow\bar{\Phi}^{II}_{20}$\\ \hline
$-4$&$\Phi^{II}_{18}$&$bF_{119}$, $bF_{181}$, $(bF_{109},bF_{110})\rightarrow \Phi^{II}_{18}$\\ \hline
Non-chiral  III\\ \hline
$0$&$\Phi^{III}_{1}$& \begin{tabular}{c} $bF_5\equiv \bar{\Phi}^{III}_{1},  (bF_{56},bF_{68})\rightarrow \Phi^{III}_{1}, $\\
$bF_{176}, bF_{156}\rightarrow \Phi^{III}_{1},\  bF_{162},bF_{138},bF_{101}\rightarrow \bar{\Phi}^{III}_{1}$\end{tabular}\\ \hline
$0$&$\Phi^{III}_{24}$& $bF_{131}\rightarrow \Phi^{III}_{24}$,  $bF_{167}\rightarrow \bar{\Phi}^{III}_{24}$\\ \hline
\end{tabular}
\end{center}
\label{singlets1}
\end{table}

The field redefinitions in Appendix \ref{redefinitions} usually involve blow--up modes from different fixed sets than those of 
the orbifold twisted fields. Although it also occurs that only the local blow--up modes take part in the redefinition. Due to the topology of the $T^6/\mathbb{Z}_{6II}$ orbifold and its resolution, this was expectable. 



\section{Anomaly cancellation in 4d}

In this section we present the study of the anomaly cancelation in 4d from the orbifold perspective
and the resolution perspective. We aim to check the equivalence of the 4d anomaly cancellation in the
orbifold deformed by vevs and in the resolution. The relevant formulas for the dimensional reduction needed
to compute the resolution anomalies can be found in \citep{Blaszczyk:2011ig}. We chose a basis inside
the Cartan subalgebra of $E_8\times E_8$ such that the abelian gauge group $U(1)^8$ is explicit and we can 
express the anomaly polynomials in terms of it.  This basis is given in Appendix \ref{U1Z6}. 

In the blow--up model the $U(1)\times SU(6)^2$ anomalies cancel. We checked that the dimensionally reduced
polynomial coincides with the one computed from the supergravity 4d spectrum. Details of the anomaly polynomials are given
in the following. We explicitly write the anomaly polynomials of the orbifold (orb), blow--up (bu) and the polynomial variation
due to field redefinitions (red). We use the symbols $I_G^{\text{orb}}, I_G^{\text{red}}$ and $I_G^{\text{bu}}$ to denote the anomaly polynomial for
the gauge factors $U(1)\text{--}G^2$ with $G=SU(2),SU(3),SU(6)$.  Also we employ the notation $F_{\text{su}(n)}$ to denote the field strength
taking values in the adjoint of $SU(n)$. The other symbols are $I_{\text{grav}}^{\text{orb,bu,red}}$ to denote the $U(1)\text{--}\text{grav}^2$ anomalies, and $I_{\text{pure}}^{\text{orb,bu,red}}$ to denote the pure $U(1)$ anomalies. 

The dimensionally reduced anomaly polynomial on $\widehat{T^6/\mathbb{Z}_{6II}}$ is given by \citep{Blumenhagen:2005ga,Nibbelink:2009sp}
\begin{align}
\label{eq:4d_Anomaly_Polynomial_Raw}
I_{6} = \mathlarger{\int}\limits_{X} \Big\{ \frac{1}{6}\left( \tr[\mathcal{F}^{\prime} F^{\prime}] \right)^{2} \!+\! \frac{1}{4}\left(\!\tr\mathcal{F}^{\prime 2}\!-\!\frac{1}{2}\tr\mathcal{R}^{2}\!\right)\tr F^{\prime 2} \!-\!\frac{1}{8}\left(\!\tr\mathcal{F}^{\prime 2}\!-\!\frac{5}{12}\tr\mathcal{R}^{2}\!\right)\tr R^{2} \Big\} \tr [\mathcal{F}^{\prime}F^{\prime}] \!+\! (^{\prime}\rightarrow ^{\prime\prime})\,.
\end{align}
The orbifold and resolution anomalies can also be explicitly evaluated using the traditional method with the charges of the fields. As a cross
check for the resolution we use both methods. The change to the anomaly due to field redefinitions is computed 
considering the remaining mass fields after symmetry breaking and computing the traces as discussed in \citep{Blaszczyk:2011ig}.

The $U(1)$--$SU(3)^2$ anomalies are given by

\begin{eqnarray}
I_{\text{su(3)}}^{\text{orb}}&=&-\frac{52}{9} F_1\tr F^2_{\text{su}(3)},\label{su3Orb}\\
I_{\text{su(3)}}^{\text{bu}}&=&\frac{1}{2}(11 F_1 +2 F_2 - 30 F_3 + 330 F_4 + 1053 F_5 - 243 F_6 - 2087 F_7 - 594 F_8)\tr F^2_{\text{su}(3)},\nonumber\\
I_{\text{su(3)}}^{\text{red}}&=&\frac{1}{6} \left(\frac{203}{3} F_1 + 6 F_2 - 90 F_3 + 990 F_4 + 3159 F_5 - 
   729 F_6- 6261 F_7 - 1782 F_8\right)\tr F^2_{\text{su}(3)}.\nonumber\end{eqnarray}
It is clear from (\ref{su3Orb})  that in the orbifold the anomalies are universal, with the unique axion canceling the $U(1)_1$--$SU(3)^2$ anomaly, whereas
in the blow--up all the $U(1)$ become anomalous. The $U(1)$--$SU(2)^2$ anomalies have  an identical structure:

\begin{eqnarray}
I_{\text{su(2)}}^{\text{orb}}&=&-\frac{52}{9} F_1\tr F^2_{\text{su}(2)},\label{su2Orb}\\
I_{\text{su(2)}}^{\text{bu}}&=&\frac{1}{2}(11 F_1 +2 F_2 - 30 F_3 + 330 F_4 + 1053 F_5 - 243 F_6 - 2087 F_7 - 594 F_8)\tr F^2_{\text{su}(2)},\nonumber\\
I_{\text{su(2)}}^{\text{red}}&=&\frac{1}{6} \left(\frac{203}{3} F_1 + 6 F_2 - 90 F_3 + 990 F_4 + 3159 F_5 - 
   729 F_6- 6261 F_7 - 1782 F_8\right)\tr F^2_{\text{su}(2)}.\nonumber
\end{eqnarray}

On the  other hand the $U(1)$--$SU(6)^2$ anomaly has a very particular structure:
\begin{eqnarray}
I_{\text{su(6)}}^{\text{orb}}&=&-\frac{52}{9} F_1 \tr F^2_{\text{su}(6)},\\
I_{\text{su(6)}}^{\text{red}}&=&\frac{52}{9} F_1 \tr F^2_{\text{su}(6)},\\
I_{\text{su(6)}}^{\text{bu}}&=&0.
\end{eqnarray}
As expected, in the orbifold the anomaly is universal, and in blow--up it turns out to be zero. 
The gravitational anomalies are given by
\begin{eqnarray}
I_{\text{grav}}^{\text{orb}}&=&\frac{52}{9} F_1\tr R^2,\\
I_{\text{grav}}^{\text{bu}}&=&-\frac{1}{12}( 23 F_1 + 7 F_2 - 119 F_3 + 1439 F_4 + 3946 F_5 + 6 (-57 F_6 - 967 F_7 + F_8))\tr R^2,\nonumber \\
I_{\text{grav}}^{\text{red}}&=&-\frac{1}{36} (277 F_1 + 3 (7 F_2 - 119 F_3 + 1439 F_4 + 3946 F_5 + 
      6 (-57 F_6 - 967 F_7 + F_8)))\tr R^2.\nonumber
\end{eqnarray}
The pure $U(1)$ anomalies have also a universal character in the blow--up:
\begin{eqnarray}
I_{pure}^{\text{orb}}&=&\frac{1}{6} \left(-\frac{10816}{27} F_1^3 - \frac{260}{9} F_1 F_2^2 - \frac{13520}{3} F_1 F_2^2 - 
\frac{1879280}{3} F_1 F_4^2 - \frac{17809792}{3} F_1 F_5^2 \right)\nonumber
\\&-&\frac{1}{6} \left( \frac{40616576}{3} F_1 F_6^2 -\frac{59672080}{3} F_1 F_7^2 -7830784 F_1 F_8^2\right).
\end{eqnarray}
On the blow--up the expression is much longer, so we refrain from writing it. It is important 
to mention the fact that compactifying in the blow--up all the $U(1)$s become anomalous.

We don't need the explicit field redefinitions obtained in order to match the anomalies in the supergravity
and in the orbifold deformed by vevs. Any map that identifies the orbifold and blow--up massless spectrum gives the
same $I^{\text{red}}$. Nevertheless, in Appendix \ref{redefinitions} we give a list of the redefined orbifold fields 
and one of the many possible redefinitions that can be used to perform the considered map.

\paragraph{Blow--up modes and non--universal axions}

Let us explore how the orbifold axion and the blow--up modes are related to the blow--up universal-- and non--universal axions.
As in the $T^6/\mathbb{Z}_7$ study \citep{Blaszczyk:2011ig} we want to determine if the local blow--up modes can be interpreted as the non--universal axions.
For that purpose we write the  anomaly change due to redefinitions as $I^{\text{red}}=\sum_r q^r_I F^I X^{\text{red}}_{4,r}$ i.e. as a factorization that can be canceled by a counterterm of blow--up modes. Then, the anomaly polynomial in the resolved space can be written as
\begin{eqnarray}
I_6=F_1X^{\text{orb}}_4+\sum_r q^r_I F^I X^{\text{red}}_{4,r}=X^{\text{uni}}_2X^{\text{uni}}_4+\sum X^r_2 X^r_4.
\end{eqnarray}
To describe the factorization we use the formulas for $X^{\text{uni}}_2,\, X^{\text{uni}}_4,\, X^r_2$ and $X^r_4$ obtained in \citep{Blaszczyk:2011ig}. To determine
how the anomaly change due to redefinitions factorizes we employ the ansatz
\begin{equation}
X^{\text{red}}_{4,r}=-\frac{1}{12} (c_r X^{\text{uni}}_{4,r}+d_r X^r_4),
\end{equation}
in which the $-1/12$ is introduced in order to simplify the normalization.  In Appendix \ref{axionscoef} we give the solutions for $c_r$ and $d_r$. Our results identify the blow--up modes $\tau_r$ as the non--universal axions $\beta_r$. The blow--up universal axion $a^{\text{uni}}$ is given as a mixture of the blow--up modes and the orbifold axion $a^{\text{orb}}$. This can be seen in the
following relations 
\begin{eqnarray}
a^{\text{uni}}&=&-\frac{1}{12}(a^{\text{orb}}+\sum_r c_r\tau_r),\\
\beta_r&=&-\frac{1}{12}d_r \tau_r.
\end{eqnarray}
The proportionality factor $-1/12 d_r$ can be chosen to be universal. It is $1/6$ for all the blow--up modes which are right--handed and $-1/6$ for the three blow--up modes which are left--handed. This result  agrees exactly with the one encountered in \citep{Blaszczyk:2011ig} for the $T^6/\mathbb{Z}_7$ orbifold. In the appendix it can also be seen that the universal blow--up axion  receives contributions from the unique orbifold axion $a^{\text{orb}}$ and the blow--up modes. This one--loop computation provides a direct identification between the orbifold resolution and the deformed orbifold with vevs of twisted fields turned on. 


\section{Conclusions and Outlook}

Our work explores deformations in heterotic orbifold compactifications  by vevs of twisted fields which can be identified with compactifications of the heterotic string on Calabi--Yau manifolds. The identification we study fulfills the following requirements. First, the blow--up modes are identified with twisted states, then the massless spectra map to each other and finally the 4d anomaly cancellation matches on both sides. The study focuses on the heterotic orbifold $T^6/\mathbb{Z}_{6II}$ and its resolution $\widehat{T^6/\mathbb{Z}_{6II}}$.   This orbifold model belongs to the MSSM  Mini--landscape which is a phenomenologically fertile region of the heterotic string compactifications. The model has the greatest complexity encountered in heterotic orbifolds. There are fixed points with $\mathbb{C}^3/\mathbb{Z}_{6II}$ singularities and fixed tori with $\mathbb{C}^2/\mathbb{Z}_{3}$ or $\mathbb{C}^2/\mathbb{Z}_{2}$ local singularities.  Part of the geometric complexity of the model is due to the singularity $\mathbb{C}^3/\mathbb{Z}_{6II}$  because it can be resolved in five different ways, leading to $\sim 5^{12}$ possibilities to resolve the compact variety.  A further complexity comes from the existence of orbifold brother models whose gauge embedding differs by lattice vectors of $\Gamma_8\times \Gamma_8$.  This creates an ambiguity in the identification of the blow--up geometry and the corresponding orbifold deformation, because the identification of the vectors determining the flux with the local orbifold shifts is only up to lattice vectors. 
 
We scanned over the Mini--landscape models, restricting the search to the ones in which all fixed sets
support chiral matter multiplets. Then, for a given orbifold model, we explored multiple resolutions.  
A technical observation is that  the Bianchi Identities are easier to fulfill by fixing the triangulation of all the local resolutions
to be the same. For triangulation B in all the fixed points resolutions, we identified many sets of twisted fields which can 
play the role of blow--up modes. Taking one of those resolutions we succeed to perform field redefinitions that  reproduce
the chiral asymmetry of the supergravity on the Calabi--Yau manifold. We looked at the masses generated by Yukawa couplings
to blow--up modes and we found that they strongly  restrict  the allowed redefinitions.  We obtained a match between the massless
spectrum of the supergravity on the blow--up and the one of the deformed orbifold. We found many equivalent
redefinitions which lead to the same identification of the orbifold spectrum with the blow--up spectrum.


One of our findings is that the local index theorem seems not applicable. That can be expected due to the presence of fixed tori,
and the absence of some exceptional divisors on the triple intersections.  Another observation is that field redefinitions
involve also non local blow--up modes. This can be expected from the fact that every two fixed sets of $T^6/\mathbb{Z}_{6II}$
have a non-vanishing spatial overlap.  With this information at hand we carried out a detailed analysis of the anomaly cancelation mechanism. 
We computed the dimensional reduced anomaly polynomial on the blown--up orbifold $\widehat{T^6/\mathbb{Z}_{6II}}$. We also obtained the orbifold anomaly polynomial and its variation due to field redefinitions and fields becoming massive on the blow--up geometry. The anomaly cancellation in 4d is inherited from the 10d cancellation. This is checked by obtaining the factorization of the 4d polynomial on $\widehat{T^6/\mathbb{Z}_{6II}}$. We were able to factorize the variation of the orbifold anomaly polynomial, and we identified the blow--up modes to be the non--universal axions
of the resolution. The universal axion on the blown--up geometry is a mixture of the orbifold--axion and the blow--up modes.  
This mixing of the axions is relevant for the interactions in blow-up. This study completes the identification of the smooth 
geometry with the deformed orbifold at the quantum level. 

Let us conclude by pointing out some problems related to this work that we would like to address in the future. It is interesting to understand the degeneracy of the identification. We would like to explore if there are stronger restrictions which could single out a bijection between a particular resolution  and a corresponding orbifold deformed by vevs.  We would like also to study how the Bianchi Identities translate into the level--matching condition for the blow--up modes. In addition, it would be interesting to study algebraic descriptions of the global Calabi--Yau manifolds with bundles, using the
understanding of the moduli space of these compactifications achieved in this work. 


Our analysis shows that to study the blow-up mechanism in detail allows us to translate the powerful computational techniques of orbifold compactification to smooth compactifications. We have shown here that even in the more complex orbifold constructions it is possible to study the orbifold--resolution transition in great detail.

\section*{Acknowledgements}  
We would like to thank M.~Blaszczyk, A.~Cabo,  S.~Groot Nibbelink, A.~Klemm, S.~Ramos S\'anchez,  F.~R\"uhle, M.~Schmitz, M.~Trapletti, P.~Vaudrevange, D.~Vieira Lopes, and I.~Zavala  for useful discussions and comments. N.G.~Cabo Bizet thanks the support of ``Centro de Aplicaciones Tecnol\'ogicas y Desarrollo Nuclear " (CEADEN,Cuba) and ``Proyecto Nacional de Ciencias B\'asicas Part\'iculas y Campos" (CITMA, Cuba).  Our work was partially supported by the SFB-Tansregio TR33 ``The Dark Universe" (Deutsche Forschungsgemeinschaft) and the European Union 7th network program ``Unification in the LHC era" (PITN-GA-2009-237920). N.G.~Cabo Bizet thanks specially the support of the program ``Unification in the LHC era" during her stay at CERN where this project was completed.

\newpage

\begin{appendix}

\section{Orbifold data}
\label{orbifoldtables}

\begin{table}[h]
\caption{Conjugacy classes of the $\mathbb{Z}_{6II}$ orbifold. In the indices $(\alpha,\beta,\gamma)$ we denote 
also by $1$ the fixed tori.}
\begin{center}
\scriptsize
\begin{longtable}{|c|c|c|c|c|} \hline
$n_{\alpha}e_{\alpha}$&$\{n_1,n_2,n_3,n_4,n_5,n_6\}$&k&$\{\alpha,\beta,\gamma\}$&F.P. coordinates\\ \hline
0&\{0, 0, 0, 0, 0, 0\} &1& \{1, 1, 1\}& \{0, 0, 0, 0, 0, 0\}\\
$e_{6} $&\{0, 0, 0, 0, 0, 1\}& 1& \{1, 1, 3\}& \{0, 0, 0, 0, 0, 1/2\}\\
$e_{5}$ & \{0, 0, 0, 0, 1, 0\}& 1&\{1, 1, 2\}& \{0, 0, 0, 0, 1/2, 0\}\\
$e_5 + e_6 $& \{0, 0, 0, 0, 1, 1\}& 1& \{1, 1, 4\}& \{0, 0, 0, 0, 1/2, 1/2\}\\
$e_3 $& \{0, 0, 1, 0, 0, 0\}& 1&\{1, 2, 1\}& \{0, 0, 2/3, 1/3, 0, 0\}\\
$e_3 + e_6$ & \{0, 0, 1, 0, 0, 1\}& 1& \{1, 2, 3\}& \{0, 0, 2/3, 1/3, 0, 1/2\}\\
$e_3 + e_5$ & \{0, 0, 1, 0, 1, 0\}& 1&\{1, 2, 2\}& \{0, 0, 2/3, 1/3, 1/2, 0\}\\
$e_3 + e_5 + e_6$ & \{0, 0, 1, 0, 1, 1\}& 1& \{1, 2, 4\}& \{0, 0, 2/3, 1/3, 1/2, 1/2\}\\
$e_3 + e_4 $&\{0, 0, 1, 1, 0, 0\}& 1& \{1, 3, 1\}& \{0, 0, 1/3, 2/3, 0, 0\}\\
$e_3 + e_4 + e_6 $& \{0, 0, 1, 1, 0, 1\}&  1&\{1, 3, 3\}& \{0, 0, 1/3, 2/3, 0, 1/2\}\\
$e_3 + e_4 + e_5 $ &\{0, 0, 1, 1, 1, 0\}& 1&\{1, 3, 2\}& \{0, 0, 1/3, 2/3, 1/2, 0\}\\
$e_3 + e_4 + e_5 + e_6 $& \{0, 0, 1, 1, 1, 1\}&  1& \{1, 3, 4\}& \{0, 0, 1/3, 2/3, 1/2, 1/2\}\\
-2 $e_2 $&\{0, -2, 0, 0, 0, 0\}& 2& \{5, 1, 1\}& \{2/3, 0, 0, 0, 0, 0\}\\
-2 $e_2$ + $e_4$ & \{0, -2, 0, 1, 0, 0\}& 2& \{5, 3, 1\}& \{2/3, 0, 1/3, 2/3, 0, 0\}\\
-2 $e_2$ + $e_3$ + $e_4$ & \{0, -2, 1, 1, 0, 0\}& 2& \{5, 2, 1\}& \{2/3, 0, 2/3, 1/3, 0, 0\}\\
 0&\{0, 0, 0, 0, 0, 0\}& 2&\{1, 1, 1\}& \{0, 0, 0, 0, 0, 0\}\\
$e_4$ &\{0, 0, 0, 1, 0, 0\}& 2&\{1, 3, 1\}& \{0, 0, 1/3, 2/3, 0, 0\}\\
$e_3 + e_4$  & \{0, 0, 1, 1, 0, 0\}& 2&\{1, 2, 1\}& \{0, 0, 2/3, 1/3, 0, 0\}\\
0&\{0, 0, 0, 0, 0, 0\}& 3& \{1, 1, 1\}& \{0, 0, 0, 0, 0, 0\}\\
$e_6 $& \{0, 0, 0, 0, 0, 1\}& 3& \{1, 1, 3\}& \{0, 0, 0, 0, 0, 1/2\} \\
$e_5$ & \{0, 0, 0, 0, 1, 0\}& 3& \{1, 1, 2\}& \{0, 0, 0, 0, 1/2, 0\}\\
$e_5 + e_6 $& \{0, 0, 0, 0, 1, 1\}& 3& \{1, 1, 4\}& \{0, 0, 0, 0, 1/2, 1/2\} \\
$e_2 $& \{0, 1, 0, 0, 0, 0\}& 3& \{4, 1, 1\}& \{0, 1/2, 0, 0, 0, 0\}\\
$e_2 + e_6$ & \{0, 1, 0, 0, 0, 1\}& 3& \{4, 1, 3\}& \{0, 1/2, 0, 0, 0, 1/2\}\\
$e_2 + e_5 $& \{0, 1, 0, 0, 1, 0\}& 3& \{4, 1, 2\}& \{0, 1/2, 0, 0, 1/2, 0\}\\
$e_2 + e_5 + e_6$ & \{0, 1, 0, 0, 1, 1\}& 3& \{4, 1, 4\}& \{0, 1/2, 0, 0, 1/2, 1/2\}\\
 0& \{0, 0, 0, 0, 0, 0\}& 4& \{1, 1, 1\}& \{0, 0, 0, 0, 0, 0\}\\
$e_3$ & \{0, 0, 1, 0, 0, 0\}& 4& \{1, 2, 1\}& \{0, 0, 2/3, 1/3, 0, 0\}\\
$e_3 + e_4$ & \{0, 0, 1, 1, 0, 0\}& 4& \{1, 3, 1\}& \{0, 0, 1/3, 2/3, 0, 0\}\\
 $e_1 + e_2$ &\{1, 1, 0, 0, 0, 0\}& 4& \{3, 1, 1\}& \{1/3, 0, 0, 0, 0, 0\}\\
 $e_1 + e_2 + e_3$ & \{1, 1, 1, 0, 0, 0\}& 4& \{3, 2, 1\}& \{1/3, 0, 2/3, 1/3, 0, 0\}\\
 $e_1 + e_2 + e_3 + e_4$ & \{1, 1, 1, 1, 0, 0\}&4& \{3, 3, 1\}& \{1/3, 0, 1/3, 2/3, 0, 0\}\\ \hline
 \end{longtable}
\end{center}
\label{conjclZ6II}
\end{table}

The orbifold Coxeter element is
\begin{eqnarray}
\theta e_1&=& 2 e_1+3 e_2,\, \theta e_2=-e_1-e_2,\,  \theta e_3=e_4,\, \theta e_4=-e_3-e_4,\\
\theta e_5&=&-e_5,\, \theta e_6=-e_6.\nonumber
\end{eqnarray}
\normalsize

\section{Field redefinitions for $T^6/\mathbb{Z}_{6II}$}
\label{redefinitions}

Here we present a sample of the field redefinitions found which perform the presented map from orbifold to blow--up states. 
The first column of the table denotes the orbifold field to be redefined $bF_{\gamma}$, the second element represents its fixed point and in the third column one can read off the redefinition. For the fields $(bF_{89}+bF_{111})_{1,2}$, $(bF_{116}+bF_{105})_{1,2}$ and $(bF_{30}+bF_{36}+bF_{125}+bF_{133})_{1,2,3,4}$, which represent mass eigenstates which are linear combinations of the terms in the sums, we write 
the redefinitions for one of the fields, instead of writing the real redefinition that is the one of the eigenstates.

\scriptsize

\begin{center}
\begin{longtable}{|c|c|c|}\hline
Field & fixed point & redefinition\\ \hline
$bF_{61}$&$(1,1,1)$&$V_{1,3,1}-V_{2,1,2} $\\
$bF_{60}$&$(1,1,1)$&$-V_{1,1,1} $\\
$bF_{62}$&$(1,1,1)$&$V_{1,3,1}-V_{2,1,2} $\\
$bF_{58}$&$(1,1,1)$&$-V_{1,1,1} $\\
$bF_{68}$&$(1,1,1)$&$-V_{1,1,1}+V_{2,3,3}+V_{4,1,3} $\\
$bF_{67}$&$(1,1,1)$&$-V_{1,2,1}+V_{4,1,1}-V_{4,3,3} $\\
$bF_{59}$&$(1,1,1)$&$2 V_{1,2,2}-V_{1,2,3}+V_{4,1,3} $\\
$bF_{65}$&$(1,1,1)$&$-V_{1,1,1} $\\
$bF_{64}$&$(1,1,1)$&$-V_{1,1,1} $\\
$bF_{66}$&$(1,1,1)$&$-V_{1,1,1} $\\
$bF_{63}$&$(1,1,1)$&$-V_{1,1,1}-V_{2,3,3}-V_{4,1,3} $\\
$bF_{42}$&$(1,1,2)$&$V_{1,1,2}+V_{2,1,3}-V_{4,3,2} $\\
$bF_{43}$&$(1,1,2)$&$V_{1,2,2}+V_{4,1,3} $\\
$bF_{49}$&$(1,1,3)$&$V_{1,3,3}-V_{2,1,2} $\\
$bF_{48}$&$(1,1,3)$&$-V_{1,1,3} $\\
$bF_{50}$&$(1,1,3)$&$V_{1,3,3}-V_{2,1,2} $\\
$bF_{46}$&$(1,1,3)$&$-V_{1,1,3} $\\
$bF_{56}$&$(1,1,3)$&$-V_{1,1,3}+V_{2,3,3}+V_{4,1,3} $\\
$bF_{53}$&$(1,1,3)$&$-V_{1,2,3}+V_{4,1,1}-V_{4,3,3} $\\
$bF_{47}$&$(1,1,3)$&$2 V_{1,2,2}-V_{1,2,3}+V_{4,1,3} $\\
$bF_{55}$&$(1,1,3)$&$-V_{1,1,3} $\\
$bF_{52}$&$(1,1,3)$&$-V_{1,1,3} $\\
$bF_{54}$&$(1,1,3)$&$-V_{1,1,3} $\\
$bF_{51}$&$(1,1,3)$&$-V_{1,1,3}-V_{2,3,3}-V_{4,1,3} $\\
$bF_{39}$&$(1,1,4)$&$V_{1,1,4}+V_{2,1,3}-V_{4,3,2} $\\
$bF_{40}$&$(1,1,4)$&$V_{1,2,4}+V_{4,1,3} $\\
$bF_{86}$&$(1,2,1)$&$-V_{1,2,1} $\\
$bF_{87}$&$(1,2,1)$&$V_{1,1,1}+V_{4,1,3} $\\
$bF_{76}$&$(1,2,2)$&$-V_{1,2,1}-V_{1,3,1}+V_{1,3,2} $\\
$bF_{79}$&$(1,2,2)$&$-V_{1,1,1}+V_{1,2,1}-V_{1,3,2} $\\
$bF_{80}$&$(1,2,2)$&$-V_{1,3,2}+V_{2,1,3}+V_{4,1,1} $\\
$bF_{78}$&$(1,2,2)$&$-V_{1,1,2}-V_{2,1,3}-V_{4,1,1} $\\
$bF_{82}$&$(1,2,2)$&$V_{1,2,2}+V_{4,3,2} $\\
$bF_{81}$&$(1,2,2)$&$-V_{1,1,1}+V_{1,2,1}-V_{1,3,2} $\\
$bF_{83}$&$(1,2,3)$&$-V_{1,2,3} $\\
$bF_{84}$&$(1,2,3)$&$V_{1,1,3}+V_{4,1,3} $\\
$bF_{69}$&$(1,2,4)$&$-V_{1,2,1}-V_{1,3,1}+V_{1,3,4} $\\
$bF_{72}$&$(1,2,4)$&$-V_{1,1,1}+V_{1,2,1}-V_{1,3,4} $\\
$bF_{73}$&$(1,2,4)$&$-V_{1,2,1}+V_{3,1,4}-V_{3,2,1} $\\
$bF_{71}$&$(1,2,4)$&$-V_{1,1,4}-V_{2,1,2}+V_{2,3,3} $\\
$bF_{74}$&$(1,2,4)$&$V_{1,2,4}-V_{2,1,2} $\\
$bF_{75}$&$(1,2,4)$&$-V_{1,1,1}+V_{1,2,1}-V_{1,3,4} $\\
$bF_{37}$&$(1,3,1)$&$V_{1,3,1}+V_{4,1,3} $\\
$bF_{36}$&$(1,3,1)$&$-V_{1,3,2}+V_{3,1,2}-V_{3,2,1} $\\
 $bF_{38}$&$(1,3,1)$&$-V_{1,3,1} $\\
$bF_{33}$&$(1,3,1)$&$V_{1,3,1}+V_{4,1,3} $\\
$bF_{35}$&$(1,3,1)$&$V_{1,3,1}-V_{2,3,3} $\\
$bF_{24}$&$(1,3,2)$&$-V_{1,3,2}$\\
$bF_{23}$&$(1,3,2)$&$V_{1,3,2}-V_{2,1,3} $\\
$bF_{20}$&$(1,3,2)$&$-V_{1,3,2} $\\
$bF_{26}$&$(1,3,2)$&$V_{1,2,2}-V_{1,2,3}-V_{1,3,3} $\\
$bF_{25}$&$(1,3,2)$&$-V_{1,3,2} $\\
$bF_{21}$&$(1,3,2)$&$-V_{1,2,2}-V_{2,3,2}-V_{4,1,3} $\\
$bF_{31}$&$(1,3,3)$&$V_{1,3,3}+V_{4,1,3} $\\
$bF_{30}$&$(1,3,3)$&$V_{1,3,3}+V_{4,1,3} $\\
$bF_{32}$&$(1,3,3)$&$-V_{1,3,3} $\\
$bF_{27}$&$(1,3,3)$&$V_{1,3,3}+V_{4,1,3} $\\
$bF_{29}$&$(1,3,3)$&$V_{1,3,3}-V_{2,3,3} $\\
$bF_{17}$&$(1,3,4)$&$-V_{1,2,1}+V_{1,2,4}-V_{1,3,1} $\\
$bF_{16}$&$(1,3,4)$&$-V_{1,3,4} $\\
$bF_{13}$&$(1,3,4)$&$V_{1,3,4}-V_{2,1,3} $\\
$bF_{19}$&$(1,3,4)$&$-V_{1,2,1}+V_{1,2,4}-V_{1,3,1} $\\
$bF_{18}$&$(1,3,4)$&$-V_{1,3,4} $\\
$bF_{14}$&$(1,3,4)$&$-V_{1,2,4}-V_{2,3,2}-V_{4,1,3} $\\
$bF_{114}$&$(2,1,1)$&$-V_{1,2,2}+V_{3,1,2}-V_{4,1,3} $\\
$bF_{111}$&$(2,1,2)$&$-V_{2,1,1}+V_{2,1,2}-V_{2,3,3} $\\
$bF_{112}$&$(2,1,2)$&$V_{1,2,2}-V_{3,1,2} $\\
$bF_{113}$&$(2,1,2)$&$-V_{2,1,2} $\\
$bF_{117}$&$(2,1,3)$&$V_{1,3,3}-V_{3,1,3} $\\
$bF_{116}$&$(2,1,3)$&$V_{1,3,2}-V_{3,1,2} $\\
$bF_{119}$&$(2,1,3)$&$-V_{2,1,3} $\\
$bF_{102}$&$(2,3,1)$&$V_{4,3,1} $\\
$bF_{100}$&$(2,3,1)$&$-V_{2,3,1} $\\
$bF_{99}$&$(2,3,1)$&$-V_{2,1,1} $\\
$bF_{101}$&$(2,3,1)$&$-V_{1,3,2}-V_{3,1,2}-V_{4,3,2} $\\
$bF_{98}$&$(2,3,1)$&$-2 V_{1,3,4}-V_{2,3,2}-V_{4,3,1} $\\
$bF_{89}$&$(2,3,2)$&$-V_{2,3,2}$\\
$bF_{93}$&$(2,3,2)$&$V_{1,1,2}+V_{1,2,2}+V_{2,3,1} $\\
$bF_{94}$&$(2,3,2)$&$V_{1,1,2}+V_{1,2,2}+V_{2,1,1} $\\
$bF_{92}$&$(2,3,2)$&$V_{1,2,2}-V_{3,1,2} $\\
$bF_{91}$&$(2,3,2)$&$-2 V_{1,3,1}-V_{2,3,2}-V_{4,1,3} $\\
$bF_{96}$&$(2,3,2)$&$2 V_{1,2,1}-V_{2,1,3}+V_{4,3,1} $\\
$bF_{95}$&$(2,3,2)$&$2 V_{1,2,1}-V_{2,1,3}+V_{4,3,1} $\\
$bF_{106}$&$(2,3,3)$&$V_{1,3,1}+V_{3,2,1} $\\
$bF_{108}$&$(2,3,3)$&$V_{1,1,2}+V_{1,3,4}+V_{2,3,1} $\\
$bF_{105}$&$(2,3,3)$&$-V_{1,1,2}-V_{1,2,2} $\\
$bF_{110}$&$(2,3,3)$&$V_{4,3,3} $\\
$bF_{109}$&$(2,3,3)$&$V_{4,3,3} $\\  
$bF_{104}$&$(2,3,3)$&$-2 V_{1,2,1}-V_{2,1,1}-V_{4,3,3} $\\
$bF_{107}$&$(2,3,3)$&$-V_{1,3,1}+V_{2,3,3}-V_{3,2,1} $\\
$bF_{182}$&$(4,1,1)$&$-V_{2,1,1}+V_{2,1,2}+V_{2,1,3} $\\
$bF_{185}$&$(4,1,1)$&$V_{1,1,1}-V_{1,2,1}+V_{2,1,3} $\\
$bF_{184}$&$(4,1,1)$&$-V_{1,1,2}-V_{1,3,2}+V_{4,1,3} $\\
$bF_{189}$&$(4,1,2)$&$-V_{4,1,2} $\\
$bF_{188}$&$(4,1,2)$&$V_{1,3,2}-V_{2,1,1}-V_{3,1,2} $\\
$bF_{190}$&$(4,1,2)$&$-V_{4,1,2} $\\
$bF_{178}$&$(4,1,3)$&$-V_{4,1,3} $\\
$bF_{181}$&$(4,1,3)$&$2 V_{1,1,2}+2 V_{1,2,2}-V_{2,1,3} $\\
$bF_{180}$&$(4,1,3)$&$-V_{4,1,3} $\\
$bF_{164}$&$(4,3,1)$&$-V_{4,3,1} $\\
$bF_{169}$&$(4,3,1)$&$V_{1,1,1}-V_{1,3,3}-V_{4,3,2} $\\
$bF_{166}$&$(4,3,1)$&$V_{2,3,1} $\\
$bF_{167}$&$(4,3,1)$&$V_{1,3,1}-V_{3,2,3}+V_{4,3,2} $\\
$bF_{168}$&$(4,3,1)$&$-V_{1,1,4}-V_{1,2,2}+V_{4,3,2} $\\
$bF_{175}$&$(4,3,2)$&$V_{2,3,2} $\\
$bF_{173}$&$(4,3,2)$&$-V_{1,1,2}-V_{1,2,4}-V_{2,3,1} $\\
$bF_{174}$&$(4,3,2)$&$V_{1,1,2}+V_{1,3,2} $\\
$bF_{172}$&$(4,3,2)$&$V_{1,3,2}-V_{2,1,1}-V_{3,1,2} $\\
$bF_{176}$&$(4,3,2)$&$-V_{2,3,1}+V_{4,1,3} $\\
$bF_{171}$&$(4,3,2)$&$V_{2,3,3}-V_{4,1,2}+V_{4,1,3} $\\
$bF_{177}$&$(4,3,2)$&$V_{2,3,2} $\\
$bF_{157}$&$(4,3,3)$&$-V_{4,1,2}+V_{4,1,3}-V_{4,3,1} $\\
$bF_{158}$&$(4,3,3)$&$-V_{4,1,3} $\\
$bF_{161}$&$(4,3,3)$&$V_{1,3,1}-V_{3,1,1}+V_{4,1,3} $\\
$bF_{162}$&$(4,3,3)$&$-V_{4,1,3} $\\
$bF_{156}$&$(4,3,3)$&$-V_{1,1,1}+V_{1,3,1}+V_{2,3,2} $\\
$bF_{160}$&$(4,3,3)$&$-V_{4,1,3} $\\
$bF_{163}$&$(4,3,3)$&$V_{1,3,1}-V_{2,3,3}-V_{3,1,1} $\\
$bF_{151}$&$(3,1,2)$&$V_{1,3,2}-V_{4,1,3} $\\
$bF_{152}$&$(3,1,2)$&$V_{3,1,2} $\\
$bF_{148}$&$(3,1,4)$&$-V_{1,3,4}+V_{2,1,3}-V_{4,1,3} $\\
$bF_{149}$&$(3,1,4)$&$V_{3,1,4} $\\
$bF_{143}$&$(3,2,1)$&$-V_{3,2,1} $\\
$bF_{142}$&$(3,2,1)$&$V_{3,2,1} $\\
$bF_{145}$&$(3,2,1)$&$-V_{3,2,1} $\\
$bF_{144}$&$(3,2,1)$&$-V_{1,3,1}-V_{2,1,3} $\\
$bF_{146}$&$(3,2,1)$&$-V_{1,3,1}+V_{2,3,3}-V_{4,1,3} $\\
$bF_{132}$&$(3,2,2)$&$-V_{1,1,2}-V_{2,3,1} $\\
$bF_{135}$&$(3,2,2)$&$-V_{1,2,2}-V_{2,3,2} $\\
$bF_{133}$&$(3,2,2)$&$V_{3,2,2} $\\
$bF_{129}$&$(3,2,2)$&$-V_{3,2,2} $\\
$bF_{131}$&$(3,2,2)$&$V_{1,2,2}+V_{2,3,2} $\\
$bF_{128}$&$(3,2,2)$&$V_{1,1,2}+V_{1,2,1}+V_{1,3,1} $\\
$bF_{130}$&$(3,2,2)$&$V_{1,1,2}+V_{2,3,1} $\\
$bF_{137}$&$(3,2,3)$&$-V_{3,2,3} $\\
$bF_{136}$&$(3,2,3)$&$V_{3,2,3} $\\
$bF_{138}$&$(3,2,3)$&$-V_{1,1,2}-V_{1,2,3}-V_{1,3,2} $\\
$bF_{139}$&$(3,2,3)$&$-V_{3,2,3} $\\
$bF_{140}$&$(3,2,3)$&$-V_{1,3,3}+V_{2,3,3}-V_{4,1,3} $\\
$bF_{124}$&$(3,2,4)$&$-V_{1,3,4}+V_{2,1,3}-V_{4,1,3} $\\
$bF_{127}$&$(3,2,4)$&$-V_{1,2,4}-V_{2,3,2} $\\
$bF_{125}$&$(3,2,4)$&$V_{3,2,4} $\\
$bF_{121}$&$(3,2,4)$&$-V_{3,2,4} $\\
$bF_{120}$&$(3,2,4)$&$-V_{2,3,3}+V_{3,2,4}-V_{4,1,3} $\\
$bF_{122}$&$(3,2,4)$&$V_{1,1,4}+V_{2,3,1} $\\
$bF_{123}$&$(3,2,4)$&$V_{1,2,4}-V_{2,1,2}+V_{4,1,2}$\\ \hline
\end{longtable}
\end{center}
\normalsize


\section{Blow--up spectrum}
\label{buspectrum}

\begin{longtable}{cccc}
\caption{Here we give all the blow--up states representations, together with one of its roots.}
\\ \hline
$\Phi^I$\\ \hline

  1 &(1,1) & -4 & ($\frac{1}{2}$,$\frac{1}{2}$,$\frac{1}{2}$,$\frac{1}{2}$,$\frac{1}{2}$,$\frac{1}{2}$,$\frac{1}{2}$,$\frac{1}{2}$,0,0,0,0,0,0,0,0)
\\
 2 &(1,1) & -4 & (-$\frac{1}{2}$,$\frac{1}{2}$,-$\frac{1}{2}$,$\frac{1}{2}$,$\frac{1}{2}$,$\frac{1}{2}$,$\frac{1}{2}$,$\frac{1}{2}$,0,0,0,0,0,0,0,0)
\\
 3 &(1,1) & -4 &(-1,1,0,0,0,0,0,0,0,0,0,0,0,0,0,0) \\
 4 &(3,1) & -3 & (-$\frac{1}{2}$,-$\frac{1}{2}$,-$\frac{1}{2}$,-$\frac{1}{2}$,-$\frac{1}{2}$,$\frac{1}{2}$,$\frac{1}{2}$,-$\frac{1}{2}$,0,0,0,0,0,0,0,0)
\\
 5 &(1,1) & -2 & ($\frac{1}{2}$,-$\frac{1}{2}$,-$\frac{1}{2}$,$\frac{1}{2}$,$\frac{1}{2}$,$\frac{1}{2}$,$\frac{1}{2}$,$\frac{1}{2}$,0,0,0,0,0,0,0,0)
\\
 6 &(1,1) & -2 & ($\frac{1}{2}$,$\frac{1}{2}$,-$\frac{1}{2}$,$\frac{1}{2}$,$\frac{1}{2}$,-$\frac{1}{2}$,-$\frac{1}{2}$,-$\frac{1}{2}$,0,0,0,0,0,0,0,0)
\\
 7 &(3,1) & -2 & (-$\frac{1}{2}$,$\frac{1}{2}$,$\frac{1}{2}$,-$\frac{1}{2}$,-$\frac{1}{2}$,$\frac{1}{2}$,$\frac{1}{2}$,-$\frac{1}{2}$,0,0,0,0,0,0,0,0)
\\
 8 &($\bar{3}$,1) & -2 & (-$\frac{1}{2}$,$\frac{1}{2}$,-$\frac{1}{2}$,$\frac{1}{2}$,$\frac{1}{2}$,$\frac{1}{2}$,-$\frac{1}{2}$,-$\frac{1}{2}$,0,0,0,0,0,0,0,0)
\\
 9 &(1,1) & -2 & (-$\frac{1}{2}$,$\frac{1}{2}$,-$\frac{1}{2}$,-$\frac{1}{2}$,-$\frac{1}{2}$,$\frac{1}{2}$,$\frac{1}{2}$,$\frac{1}{2}$,0,0,0,0,0,0,0,0)
\\
 10 &(1,2) & -2 & (-$\frac{1}{2}$,-$\frac{1}{2}$,-$\frac{1}{2}$,$\frac{1}{2}$,-$\frac{1}{2}$,$\frac{1}{2}$,$\frac{1}{2}$,$\frac{1}{2}$,0,0,0,0,0,0,0,0)
\\
 11 &($\bar{3}$,2) & -2 &(0,0,0,1,0,1,0,0,0,0,0,0,0,0,0,0) \\
 12 &(3,1) & -2 &(0,0,0,0,0,1,1,0,0,0,0,0,0,0,0,0) \\
 13 &(1,1) & -2 &(-1,0,1,0,0,0,0,0,0,0,0,0,0,0,0,0) \\
 14 &(1,1) & -2 &(0,1,-1,0,0,0,0,0,0,0,0,0,0,0,0,0) \\
 15 &(1,1) & -2 &(-1,0,-1,0,0,0,0,0,0,0,0,0,0,0,0,0) \\
 16 &(1,2) & -2 &(-1,0,0,-1,0,0,0,0,0,0,0,0,0,0,0,0) \\
 17 &(1,2) & -2 &(0,0,-1,-1,0,0,0,0,0,0,0,0,0,0,0,0) \\
 18 &(1,1) & -2 &(0,0,0,-1,-1,0,0,0,0,0,0,0,0,0,0,0) \\
 19 &(1,2) & -1 & (-$\frac{1}{2}$,$\frac{1}{2}$,$\frac{1}{2}$,-$\frac{1}{2}$,$\frac{1}{2}$,$\frac{1}{2}$,$\frac{1}{2}$,$\frac{1}{2}$,0,0,0,0,0,0,0,0)
\\
 20 &($\bar{3}$,2) & -1 & (-$\frac{1}{2}$,$\frac{1}{2}$,$\frac{1}{2}$,$\frac{1}{2}$,-$\frac{1}{2}$,$\frac{1}{2}$,-$\frac{1}{2}$,-$\frac{1}{2}$,0,0,0,0,0,0,0,0)
\\
 21 &(1,2) & -1 & (-$\frac{1}{2}$,$\frac{1}{2}$,-$\frac{1}{2}$,$\frac{1}{2}$,-$\frac{1}{2}$,-$\frac{1}{2}$,-$\frac{1}{2}$,-$\frac{1}{2}$,0,0,0,0,0,0,0,0)
\\
 22 &(1,1) & -1 & (-$\frac{1}{2}$,-$\frac{1}{2}$,-$\frac{1}{2}$,$\frac{1}{2}$,$\frac{1}{2}$,-$\frac{1}{2}$,-$\frac{1}{2}$,-$\frac{1}{2}$,0,0,0,0,0,0,0,0)
\\
 23 &(1,2) & -1 &(1,0,0,-1,0,0,0,0,0,0,0,0,0,0,0,0) \\
 24 &(1,1) & -1 &(-1,-1,0,0,0,0,0,0,0,0,0,0,0,0,0,0) \\
 25 &(3,1) & -1 &(0,0,-1,0,0,-1,0,0,0,0,0,0,0,0,0,0) \\ \hline
 $\Phi^{II}$\\ \hline
 2 &$\bar{6}$& -1 & (0,0,0,0,0,0,0,0,-$\frac{1}{2}$,$\frac{1}{2}$,-$\frac{1}{2}$,$\frac{1}{2}$,$\frac{1}{2}$,$\frac{1}{2}$,-$\frac{1}{2}$,-$\frac{1}{2}$)
\\
 3 & 1 & -2 & (0,0,0,0,0,0,0,0,-$\frac{1}{2}$,$\frac{1}{2}$,$\frac{1}{2}$,$\frac{1}{2}$,$\frac{1}{2}$,$\frac{1}{2}$,$\frac{1}{2}$,-$\frac{1}{2}$) \\
 4 & 6 & -1 & (0,0,0,0,0,0,0,0,$\frac{1}{2}$,$\frac{1}{2}$,-$\frac{1}{2}$,$\frac{1}{2}$,-$\frac{1}{2}$,-$\frac{1}{2}$,-$\frac{1}{2}$,$\frac{1}{2}$)\\
 7 & 1 & -2 & (0,0,0,0,0,0,0,0,-$\frac{1}{2}$,-$\frac{1}{2}$,$\frac{1}{2}$,$\frac{1}{2}$,$\frac{1}{2}$,$\frac{1}{2}$,-$\frac{1}{2}$,-$\frac{1}{2}$)\\
 9 &$\bar{6}$& -1 &(0,0,0,0,0,0,0,0,0,1,-1,0,0,0,0,0) \\
  13 & 6 & -2 &(0,0,0,0,0,0,0,0,0,0,0,1,0,0,1,0) \\
 14 &$\bar{6}$& -2 &(0,0,0,0,0,0,0,0,0,-1,-1,0,0,0,0,0) \\
  18 & 1 & -4 &(0,0,0,0,0,0,0,0,0,-1,0,0,0,0,1,0) \\
 19 & 6 & -1 &(0,0,0,0,0,0,0,0,0,0,0,1,0,0,-1,0) \\
 20 & 1 & -4 &(0,0,0,0,0,0,0,0,0,-1,0,0,0,0,-1,0)\\ \hline
 $\Phi^{III}$\\ \hline
 1 &(1,1) & 0 & (-$\frac{1}{2}$,-$\frac{1}{2}$,$\frac{1}{2}$,$\frac{1}{2}$,$\frac{1}{2}$,$\frac{1}{2}$,$\frac{1}{2}$,$\frac{1}{2}$,0,0,0,0,0,0,0,0)
\\
 13 &(3,2) & 0 & (-$\frac{1}{2}$,$\frac{1}{2}$,-$\frac{1}{2}$,$\frac{1}{2}$,-$\frac{1}{2}$,$\frac{1}{2}$,$\frac{1}{2}$,-$\frac{1}{2}$,0,0,0,0,0,0,0,0)
\\
 16 &(3,1) & 0 & (-$\frac{1}{2}$,-$\frac{1}{2}$,-$\frac{1}{2}$,$\frac{1}{2}$,$\frac{1}{2}$,$\frac{1}{2}$,$\frac{1}{2}$,-$\frac{1}{2}$,0,0,0,0,0,0,0,0)
\\
 24 &(1,1) & 0 &(0,1,1,0,0,0,0,0,0,0,0,0,0,0,0,0) \\
  29 &(1,2) & 0 &(0,1,0,-1,0,0,0,0,0,0,0,0,0,0,0,0) \\
  32 &($\bar{3}$,1) & 0 &(0,-1,0,0,0,1,0,0,0,0,0,0,0,0,0,0) \\
 \hline
\label{bustates}
\end{longtable}%
\normalsize
\section{$U(1)$ basis}
\label{U1Z6}
We start with a basis for a set of Cartan generators $H_I$ such that $\tr H_I H_J=\delta_{IJ}$. There are
8 $U(1)_k$ symmetries. Writing the generator of each of them as $G_k=\sum_I c_k^I H_I$, they
are given as 
\begin{eqnarray*}
G_1&=&(\frac{11}{6}, \frac{1}{2}, -\frac{1}{2}, -\frac{1}{2}, -\frac{1}{2}, -\frac{1}{2}, -\frac{1}{2}, -\frac{1}{2}, \frac{1}{2}, -\frac{13}{6}, -\frac{1}{2}, -\frac{1}{2}, -\frac{1}{2}, -\frac{1}{2}, \frac{1}{2}, \frac{1}{2}),\\
G_2&=&(0, 0, 0, \frac{1}{2}, \frac{1}{2}, -\frac{1}{3}, -\frac{1}{3}, -\frac{1}{3}, 0, 0, 0, 0, 0, 0, 0, 0),\\
G_3&=&(-3, 11, 0, 0, 0, 0, 0, 0, 0, 0, 0, 0, 0, 0, 0, 0),\\
G_4&=&(33, 9, 130, 0, 0, 0, 0, 0, 0, 0, 0, 0, 0, 0, 0, 0),\\
G_5&=&(286, 78, -78, 0, 0, 0, 0, 0, 0, 278, 0, 0, 0, 0, 0, 0),\\
G_6&=&(-66, -18, 18, 0, 0, 0, 0, 0, 0, 78, 0, 0, 0, 0, 616, 0),\\
G_7&=&(165, 45, -45, 317, 317, 317, 317, 317, 0, -195, 0, 0, 0, 0, 45, 0),\\
G_8&=&(-99, -27, 27, 27, 27, 27, 27, 27, 181, 117, -181, -181, -181, -181, -27, 181).
\end{eqnarray*}
Every entry $I$ in the vector $G_k$ represents the coefficient $c_k^I$. The generator $G_1$ is the
generator of the anomalous $U(1)_1$.

\section{Axions in blow--up versus orbifold axions}
\label{axionscoef}
Here we give the solutions for the coefficients $c_r$ and $d_r$ relating the orbifold
axion $a^{\text{orb}}$ and the blow--up modes $\tau_r$ with the universal $a^{\text{uni}}$ and non--universal axions 
$\beta_r$ in the resolution.
The relations are
\begin{eqnarray}
a^{\text{uni}}&=&-\frac{1}{12}(a^{\text{orb}}+\sum_r c_r\tau_r)\\
\beta_r&=&-\frac{1}{12}d_r \tau_r.
\end{eqnarray}

The following set correspond to solutions such that $I^{\text{red}}$ is factorizable and therefore can be canceled by a counterterm: 
\scriptsize
\begin{longtable}{c}
$c_1= d_{19}+\frac{45}{4} (-4-d_{28})+\frac{1}{4} (4-4 c_3+16 c_{17}+16 c_{19}-8 c_{20}-20 c_{21}+10 c_{25}$\\$
+12 c_{26}+10 c_{27}+12 c_{28}-2 c_{29}+12 c_{30}-2 c_{31}+12 c_{32}+4 d_{20}+4 d_{21}+4 d_{22}$\\$
+4 d_{23}+4 d_{24}+45 d_{28}+4 d_{29}+4 d_{30}+4 d_{31}+4 d_{32},$\\$
c_2=1-c_4+4 c_{17}+5 c_{19}-2 c_{20}-5 c_{21}+3 c_{25}+2 c_{26}+3 c_{27}+2 c_{28}-c_{29}$\\$
+2 c_{30}-c_{31}+2 c_{32}+d_{19}+d_{20}+d_{21}+d_{22}+d_{23}+d_{24}+\frac{21}{2}(-4-d_{28})$\\$
+\frac{21 d_{28}}{2}+d_{29}+d_{30}+d_{31}+d_{32},$\\$
c_9= -1-c_5-c_7-c_{11}-4 c_{17}-4 c_{19}+2 c_{20}+3 c_{21}-\frac{5 c_{25}}{2}-3c_{26}-\frac{5 c_{27}}{2}$\\$
-3 c_{28}+\frac{c_{29}}{2}-3 c_{30}+\frac{c_{31}}{2}-3 c_{32}-d_{19}-d_{20}-d_{21}-d_{22}-d_{23}-d_{24}-\frac{51}{4} (-4-d_{28})$\\$
-\frac{51d_{28}}{4}-d_{29}-d_{30}-d_{31}-d_{32}, $\\$
c_{10}= 1-c_6-c_8-c_{12}+4 c_{17}+5 c_{19}-2 c_{20}-5 c_{21}+3 c_{25}+3 c_{26}+3 c_{27}+3 c_{28}$\\$
-c_{29}+3c_{30}-c_{31}+3 c_{32}+d_{19}+d_{20}+d_{21}+d_{22}+d_{23}+d_{24}+\frac{27}{2} (-4-d_{28})$\\$
+\frac{27 d_{28}}{2}+d_{29}+d_{30}+d_{31}+d_{32}, $\\$
c_{13}=-2+c_6+c_8-7 c_{17}-8 c_{19}+3 c_{20}+9 c_{21}-6 c_{25}-5 c_{26}-6 c_{27}-5 c_{28}$\\$
+2 c_{29}-5 c_{30}+2 c_{31}-5 c_{32}-2 d_{19}-2 d_{20}-2 d_{21}-2d_{22}-2 d_{23}-2 d_{24}$\\$
-24 (-4-d_{28})-24 d_{28}-2 d_{29}-2 d_{30}-2 d_{31}-2 d_{32},$\\$
c_{14}= -1-c_5-c_7-3 c_{17}-4 c_{19}+c_{20}+3c_{21}-c_{23}-2 c_{25}-2 c_{26}-2 c_{27}-2 c_{28}$\\$
-2 c_{30}-2 c_{32}-d_{19}-d_{20}-d_{21}-d_{22}-d_{23}-d_{24}-\frac{15}{2} (-4-d_{28})$\\$
-\frac{15 d_{28}}{2}-d_{29}-d_{30}-d_{31}-d_{32},$\\$
c_{15}= -1+c_5+c_6+c_7+c_8-4 c_{17}-4 c_{19}+2 c_{20}+5 c_{21}-c_{24}-3 c_{25}-2 c_{26}$\\$
-3 c_{27}-2 c_{28}+c_{29}-2c_{30}+c_{31}-2 c_{32}-d_{19}-d_{20}-d_{21}-d_{22}-d_{23}-d_{24}$\\$
-\frac{63}{4} (-4-d_{28})-\frac{63 d_{28}}{4}-d_{29}-d_{30}-d_{31}-d_{32}, $\\$
c_{16}=1-c_6-c_8+4 c_{17}+4 c_{19}-2 c_{20}-5 c_{21}-c_{22}+3 c_{25}+3 c_{26}+3 c_{27}+3 c_{28}$\\$
-c_{29}+3 c_{30}-c_{31}+3 c_{32}+d_{19}+d_{20}+d_{21}+d_{22}+d_{23}+d_{24}$\\$
+\frac{69}{4}(-4-d_{28})+\frac{69 d_{28}}{4}+d_{29}+d_{30}+d_{31}+d_{32}, $\\$
c_{18}= -c_6-c_8+c_{17}+c_{19}-c_{20}-c_{21}+c_{25}+c_{27}-c_{29}-c_{31}$\\$
-\frac{3}{4}(-4-d_{28})-\frac{3 d_{28}}{4}$,\\
$d_1= -4-d_3,$\\
$d_2= -4-d_4,$\\
$d_5= -4-d_7,$\\
$d_6= -4-d_8,$\\
$d_9= -4-d_{11},$\\
$d_{10}= -4-d_{12},$\\
$d_{13}=-2,$\\
$d_{14}= 2,$\\
$d_{15}= 2,$\\
$d_{16}= -2,$\\
$d_{17}= -2,$\\
$d_{18}=2,$\\$
d_{25}= -4-d_{27},$\\
$d_{26}= -4-d_{28}$.
\end{longtable}
\end{appendix}

\bibliographystyle{utphys}
{\small
\bibliography{bibliography1}{}

\providecommand{\href}[2]{#2}\begingroup\raggedright\begin{thebibliography}{10}

\bibitem{Lebedev:2006kn}
O.~Lebedev, H.~P. Nilles, S.~Raby, S.~Ramos-Sanchez, M.~Ratz, {\em et~al.},
  ``{A Mini-landscape of exact MSSM spectra in heterotic orbifolds},''
  \href{http://dx.doi.org/10.1016/j.physletb.2006.12.012}{{\em Phys.Lett.}
  {\bfseries B645} (2007) 88--94},
\href{http://arxiv.org/abs/0611095}{{\ttfamily arXiv:0611095 [hep-th]}}.

\bibitem{Forste:2004ie}
S.~Forste, H.~P. Nilles, P.~K.~S. Vaudrevange, and A.~Wingerter, ``{Heterotic
  brane world},'' \href{http://dx.doi.org/10.1103/PhysRevD.70.106008}{{\em
  Phys. Rev.} {\bfseries D70} (2004) 106008},
\href{http://arxiv.org/abs/0406208}{{\ttfamily arXiv:0406208}}.

\bibitem{Kobayashi:2004ya}
T.~Kobayashi, S.~Raby, and R.-J. Zhang, ``{Searching for realistic 4d string
  models with a Pati-Salam symmetry: Orbifold grand unified theories from
  heterotic string compactification on a Z(6) orbifold},''
  \href{http://dx.doi.org/10.1016/j.nuclphysb.2004.10.035}{{\em Nucl. Phys.}
  {\bfseries B704} (2005) 3--55},
\href{http://arxiv.org/abs/hep-ph/0409098}{{\ttfamily arXiv:hep-ph/0409098}}.

\bibitem{Buchmuller:2004hv}
W.~Buchmuller, K.~Hamaguchi, O.~Lebedev, and M.~Ratz, ``{Dual models of gauge
  unification in various dimensions},''
  \href{http://dx.doi.org/10.1016/j.nuclphysb.2005.01.038}{{\em Nucl. Phys.}
  {\bfseries B712} (2005) 139--156},
\href{http://arxiv.org/abs/hep-ph/0412318}{{\ttfamily arXiv:hep-ph/0412318}}.

\bibitem{Buchmuller:2005jr}
W.~Buchmuller, K.~Hamaguchi, O.~Lebedev, and M.~Ratz, ``{Supersymmetric
  standard model from the heterotic string},''
  \href{http://dx.doi.org/10.1103/PhysRevLett.96.121602}{{\em Phys. Rev. Lett.}
  {\bfseries 96} (2006) 121602},
\href{http://arxiv.org/abs/hep-ph/0511035}{{\ttfamily arXiv:hep-ph/0511035}}.

\bibitem{Nilles:2009yd}
H.~P. Nilles, S.~Ramos-Sanchez, and P.~K.~S. Vaudrevange, ``{Local Grand
  Unification and String Theory},''
  \href{http://dx.doi.org/10.1063/1.3327561}{{\em AIP Conf. Proc.} {\bfseries
  1200} (2010) 226--234},
\href{http://arxiv.org/abs/0909.3948}{{\ttfamily arXiv:0909.3948 [hep-th]}}.

\bibitem{Nilles:2008gq}
H.~P. Nilles, S.~Ramos-Sanchez, M.~Ratz, and P.~K.~S. Vaudrevange, ``{From
  strings to the MSSM},''
  \href{http://dx.doi.org/10.1140/epjc/s10052-008-0740-1}{{\em Eur. Phys. J.}
  {\bfseries C59} (2009) 249--267},
\href{http://arxiv.org/abs/0806.3905}{{\ttfamily arXiv:0806.3905 [hep-th]}}.

\bibitem{Buchmuller:2005sh}
W.~Buchmuller, K.~Hamaguchi, O.~Lebedev, and M.~Ratz, ``{Local grand
  unification},''
\href{http://arxiv.org/abs/hep-ph/0512326}{{\ttfamily arXiv:hep-ph/0512326}}.

\bibitem{Kobayashi:2006wq}
T.~Kobayashi, H.~P. Nilles, F.~Ploger, S.~Raby, and M.~Ratz, ``{Stringy origin
  of non-Abelian discrete flavor symmetries},''
  \href{http://dx.doi.org/10.1016/j.nuclphysb.2007.01.018}{{\em Nucl. Phys.}
  {\bfseries B768} (2007) 135--156},
\href{http://arxiv.org/abs/hep-ph/0611020}{{\ttfamily arXiv:hep-ph/0611020}}.

\bibitem{CKMPZS}
N.~G. Cabo~Bizet, T.~Kobayashi, D.~K. Mayorga Pe\~na, S.~L. Parameswaran,
  M.~Schmitz, {\em et~al.}, ``{R-charge Conservation and More in Factorizable
  and Non-Factorizable Orbifolds},''
\href{http://arxiv.org/abs/1301.2322}{{\ttfamily arXiv:1301.2322 [hep-th]}}.

\bibitem{Kappl:2008ie}
R.~Kappl {\em et~al.}, ``{Large hierarchies from approximate R symmetries},''
  \href{http://dx.doi.org/10.1103/PhysRevLett.102.121602}{{\em Phys. Rev.
  Lett.} {\bfseries 102} (2009) 121602},
\href{http://arxiv.org/abs/0812.2120}{{\ttfamily arXiv:0812.2120 [hep-th]}}.

\bibitem{Forste:2010pf}
S.~Forste, H.~P. Nilles, S.~Ramos-Sanchez, and P.~K.~S. Vaudrevange, ``{Proton
  Hexality in Local Grand Unification},''
  \href{http://dx.doi.org/10.1016/j.physletb.2010.08.057}{{\em Phys. Lett.}
  {\bfseries B693} (2010) 386--392},
\href{http://arxiv.org/abs/1007.3915}{{\ttfamily arXiv:1007.3915 [hep-ph]}}.

\bibitem{Lee:2010gv}
H.~M. Lee {\em et~al.}, ``{A unique $Z_4^R$ symmetry for the MSSM},''
  \href{http://dx.doi.org/10.1016/j.physletb.2010.10.038}{{\em Phys. Lett.}
  {\bfseries B694} (2011) 491--495},
\href{http://arxiv.org/abs/1009.0905}{{\ttfamily arXiv:1009.0905 [hep-ph]}}.

\bibitem{Ko:2007dz}
P.~Ko, T.~Kobayashi, J.-h. Park, and S.~Raby, ``{String-derived D4 flavor
  symmetry and phenomenological implications},''
  \href{http://dx.doi.org/10.1103/PhysRevD.76.035005}{{\em Phys. Rev.}
  {\bfseries D76} (2007) 035005},
\href{http://arxiv.org/abs/0704.2807}{{\ttfamily arXiv:0704.2807 [hep-ph]}}.

\bibitem{Casas:1992mk}
J.~A. Casas and C.~Munoz, ``{A Natural Solution to the MU Problem},''
  \href{http://dx.doi.org/10.1016/0370-2693(93)90081-R}{{\em Phys. Lett.}
  {\bfseries B306} (1993) 288--294},
\href{http://arxiv.org/abs/hep-ph/9302227}{{\ttfamily arXiv:hep-ph/9302227}}.

\bibitem{Antoniadis:1994hg}
I.~Antoniadis, E.~Gava, K.~S. Narain, and T.~R. Taylor, ``{Effective mu term in
  superstring theory},''
  \href{http://dx.doi.org/10.1016/0550-3213(94)90599-1}{{\em Nucl. Phys.}
  {\bfseries B432} (1994) 187--204},
\href{http://arxiv.org/abs/9405024}{{\ttfamily arXiv:9405024}}.

\bibitem{Lebedev:2007hv}
O.~Lebedev {\em et~al.}, ``{The Heterotic Road to the MSSM with R parity},''
  \href{http://dx.doi.org/10.1103/PhysRevD.77.046013}{{\em Phys. Rev.}
  {\bfseries D77} (2008) 046013},
\href{http://arxiv.org/abs/0708.2691}{{\ttfamily arXiv:0708.2691 [hep-th]}}.

\bibitem{Hamidi:1986vh}
S.~Hamidi and C.~Vafa, ``{Interactions on Orbifolds},''
\href{http://dx.doi.org/10.1016/0550-3213(87)90006-X}{{\em Nucl. Phys.}
  {\bfseries B279} (1987) 465}.

\bibitem{Candelas:1985en}
P.~Candelas, G.~T. Horowitz, A.~Strominger, and E.~Witten, ``{Vacuum
  Configurations for Superstrings},''
\href{http://dx.doi.org/10.1016/0550-3213(85)90602-9}{{\em Nucl. Phys.}
  {\bfseries B258} (1985) 46--74}.

\bibitem{Erler:1992ki}
J.~Erler and A.~Klemm, ``{Comment on the generation number in orbifold
  compactifications},'' \href{http://dx.doi.org/10.1007/BF02096954}{{\em
  Commun. Math. Phys.} {\bfseries 153} (1993) 579--604},
\href{http://arxiv.org/abs/9207111}{{\ttfamily arXiv:9207111 [hep-th]}}.

\bibitem{Aspinwall:1994ev}
P.~S. Aspinwall, ``{Resolution of orbifold singularities in string theory},''
  \href{http://arxiv.org/abs/9403123}{{\ttfamily arXiv:9403123 [hep-th]}}. To
  appear in 'Essays on Mirror Manifolds 2'.

\bibitem{Reffert:2006du}
S.~Reffert, ``{Toroidal Orbifolds: Resolutions, Orientifolds and Applications
  in String Phenomenology},''
\href{http://arxiv.org/abs/0609040}{{\ttfamily arXiv:0609040 [hep-th]}}.

\bibitem{Lust:2006zh}
D.~Lust, S.~Reffert, E.~Scheidegger, and S.~Stieberger, ``{Resolved toroidal
  orbifolds and their orientifolds},'' {\em Adv. Theor. Math. Phys.} {\bfseries
  12} (2008) 67--183,
\href{http://arxiv.org/abs/0609014}{{\ttfamily arXiv:0609014}}.

\bibitem{Nibbelink:2007rd}
S.~G. Nibbelink, M.~Trapletti, and M.~Walter, ``{Resolutions of Cn/Zn
  Orbifolds, their U(1) Bundles, and Applications to String Model Building},''
  \href{http://dx.doi.org/10.1088/1126-6708/2007/03/035}{{\em JHEP} {\bfseries
  03} (2007) 035},
\href{http://arxiv.org/abs/0701227}{{\ttfamily arXiv:0701227}}.

\bibitem{Nibbelink:2007pn}
S.~G. Nibbelink, T.-W. Ha, and M.~Trapletti, ``{Toric Resolutions of Heterotic
  Orbifolds},'' \href{http://dx.doi.org/10.1103/PhysRevD.77.026002}{{\em Phys.
  Rev.} {\bfseries D77} (2008) 026002},
\href{http://arxiv.org/abs/0707.1597}{{\ttfamily arXiv:0707.1597 [hep-th]}}.

\bibitem{Witten:1981me}
E.~Witten, ``{Search for a Realistic Kaluza-Klein Theory},''
\href{http://dx.doi.org/10.1016/0550-3213(81)90021-3}{{\em Nucl.Phys.}
  {\bfseries B186} (1981) 412}.

\bibitem{Witten:1984dg}
E.~Witten, ``{Some Properties of O(32) Superstrings},''
\href{http://dx.doi.org/10.1016/0370-2693(84)90422-2}{{\em Phys. Lett.}
  {\bfseries B149} (1984) 351--356}.

\bibitem{Froggatt:1978nt}
C.~D. Froggatt and H.~B. Nielsen, ``{Hierarchy of Quark Masses, Cabibbo Angles
  and CP Violation},''
\href{http://dx.doi.org/10.1016/0550-3213(79)90316-X}{{\em Nucl. Phys.}
  {\bfseries B147} (1979) 277}.

\bibitem{Atick:1987gy}
J.~J. Atick, L.~J. Dixon, and A.~Sen, ``{String Calculation of Fayet-Iliopoulos
  d Terms in Arbitrary Supersymmetric Compactifications},''
\href{http://dx.doi.org/10.1016/0550-3213(87)90639-0}{{\em Nucl. Phys.}
  {\bfseries B292} (1987) 109--149}.

\bibitem{Dine:1987xk}
M.~Dine, N.~Seiberg, and E.~Witten, ``{Fayet-Iliopoulos Terms in String
  Theory},''
\href{http://dx.doi.org/10.1016/0550-3213(87)90395-6}{{\em Nucl. Phys.}
  {\bfseries B289} (1987) 589}.

\bibitem{Font:1988mm}
A.~Font, L.~E. Ibanez, H.~P. Nilles, and F.~Quevedo, ``{Yukawa Couplings in
  Degenerate Orbifolds: Towards a Realistic SU(3) x SU(2) x U(1)
  Superstring},''
{\em Phys. Lett.} {\bfseries 210B} (1988) 101.

\bibitem{Ludeling:2012cu}
C.~Ludeling, F.~Ruehle, and C.~Wieck, ``{Non-Universal Anomalies in Heterotic
  String Constructions},''
  \href{http://dx.doi.org/10.1103/PhysRevD.85.106010}{{\em Phys. Rev.}
  {\bfseries D85} (2012) 106010},
  \href{http://arxiv.org/abs/1203.5789}{{\ttfamily arXiv:1203.5789 [hep-th]}}.

\bibitem{Oda}
T.~Oda, {\em {Convex Bodies and Algebraic Geometry}}.
\newblock Springer--Verlag,
1988.
\newblock

\bibitem{Fulton}
W.~Fulton, {\em {Introduction to toric varieties}}.
\newblock Annals of mathematics studies ; 131\\ The William H. Roever lectures
  in geometry. Princeton University Press, 1997.

\bibitem{Hori:2003ic}
K.~Hori, S.~Katz, A.~Klemm, R.~Pandharipande, R.~Thomas, {\em et~al.},
``{Mirror symmetry},''.

\bibitem{Nibbelink:2008tv}
S.~G. Nibbelink, D.~Klevers, F.~Ploger, M.~Trapletti, and P.~K.~S. Vaudrevange,
  ``{Compact heterotic orbifolds in blow-up},''
  \href{http://dx.doi.org/10.1088/1126-6708/2008/04/060}{{\em JHEP} {\bfseries
  04} (2008) 060},
\href{http://arxiv.org/abs/0802.2809}{{\ttfamily arXiv:0802.2809 [hep-th]}}.

\bibitem{Nibbelink:2009sp}
S.~G. Nibbelink, J.~Held, F.~Ruehle, M.~Trapletti, and P.~K.~S. Vaudrevange,
  ``{Heterotic Z6-II MSSM Orbifolds in Blowup},''
  \href{http://dx.doi.org/10.1088/1126-6708/2009/03/005}{{\em JHEP} {\bfseries
  03} (2009) 005},
\href{http://arxiv.org/abs/0901.3059}{{\ttfamily arXiv:0901.3059 [hep-th]}}.

\bibitem{NibbelinkGroot:2010wm}
S.~Nibbelink~Groot, ``{Heterotic orbifold resolutions as (2,0) gauged linear
  sigma models},'' \href{http://dx.doi.org/10.1002/prop.201100002}{{\em
  Fortsch.Phys.} {\bfseries 59} (2011) 454--493},
\href{http://arxiv.org/abs/1012.3350}{{\ttfamily arXiv:1012.3350 [hep-th]}}.

\bibitem{Blaszczyk:2011ib}
M.~Blaszczyk, S.~Nibbelink~Groot, and F.~Ruehle, ``{Green-Schwarz Mechanism in
  Heterotic (2,0) Gauged Linear Sigma Models: Torsion and NS5 Branes},''
  \href{http://dx.doi.org/10.1007/JHEP08(2011)083}{{\em JHEP} {\bfseries 1108}
  (2011) 083},
\href{http://arxiv.org/abs/1107.0320}{{\ttfamily arXiv:1107.0320 [hep-th]}}.

\bibitem{Blaszczyk:2011hs}
M.~Blaszczyk, S.~Groot~Nibbelink, and F.~Ruehle, ``{Gauged Linear Sigma Models
  for toroidal orbifold resolutions},''
  \href{http://dx.doi.org/10.1007/JHEP05(2012)053}{{\em JHEP} {\bfseries 1205}
  (2012) 053},
\href{http://arxiv.org/abs/1111.5852}{{\ttfamily arXiv:1111.5852 [hep-th]}}.

\bibitem{Blaszczyk:2010db}
M.~Blaszczyk, S.~Nibbelink~Groot, F.~Ruehle, M.~Trapletti, and P.~K.
  Vaudrevange, ``{Heterotic MSSM on a Resolved Orbifold},''
  \href{http://dx.doi.org/10.1007/JHEP09(2010)065}{{\em JHEP} {\bfseries 1009}
  (2010) 065},
\href{http://arxiv.org/abs/1007.0203}{{\ttfamily arXiv:1007.0203 [hep-th]}}.

\bibitem{Buchmuller:2012mu}
W.~Buchmuller, J.~Louis, J.~Schmidt, and R.~Valandro, ``{Voisin-Borcea
  Manifolds and Heterotic Orbifold Models},''
  \href{http://dx.doi.org/10.1007/JHEP10(2012)114}{{\em JHEP} {\bfseries 1210}
  (2012) 114},
\href{http://arxiv.org/abs/1208.0704}{{\ttfamily arXiv:1208.0704 [hep-th]}}.

\bibitem{Lebedev:2008un}
O.~Lebedev, H.~P. Nilles, S.~Ramos-Sanchez, M.~Ratz, and P.~K.~S. Vaudrevange,
  ``{Heterotic mini-landscape (II): completing the search for MSSM vacua in a
  $Z_6$ orbifold},''
  \href{http://dx.doi.org/10.1016/j.physletb.2008.08.054}{{\em Phys. Lett.}
  {\bfseries B668} (2008) 331--335},
\href{http://arxiv.org/abs/0807.4384}{{\ttfamily arXiv:0807.4384 [hep-th]}}.

\bibitem{Kim:2007mt}
J.~E. Kim, J.-H. Kim, and B.~Kyae, ``{Superstring standard model from Z(12-I)
  orbifold compactification with and without exotics, and effective R-
  parity},'' \href{http://dx.doi.org/10.1088/1126-6708/2007/06/034}{{\em JHEP}
  {\bfseries 06} (2007) 034},
\href{http://arxiv.org/abs/hep-ph/0702278}{{\ttfamily arXiv:hep-ph/0702278}}.

\bibitem{Blaszczyk:2009in}
M.~Blaszczyk {\em et~al.}, ``{A Z2xZ2 standard model},''
  \href{http://dx.doi.org/10.1016/j.physletb.2009.12.036}{{\em Phys. Lett.}
  {\bfseries B683} (2010) 340--348},
\href{http://arxiv.org/abs/0911.4905}{{\ttfamily arXiv:0911.4905 [hep-th]}}.

\bibitem{Vafa:1994rv}
C.~Vafa and E.~Witten, ``{On orbifolds with discrete torsion},''
  \href{http://dx.doi.org/10.1016/0393-0440(94)00048-9}{{\em J. Geom. Phys.}
  {\bfseries 15} (1995) 189--214},
\href{http://arxiv.org/abs/9409188}{{\ttfamily arXiv:9409188}}.

\bibitem{Ploger:2007iq}
F.~Ploger, S.~Ramos-Sanchez, M.~Ratz, and P.~K.~S. Vaudrevange, ``{Mirage
  Torsion},'' {\em JHEP} {\bfseries 04} (2007) 063,
\href{http://arxiv.org/abs/0702176}{{\ttfamily arXiv:0702176}}.

\bibitem{Nibbelink:2007ew}
S.~Groot~Nibbelink, H.~P. Nilles, and M.~Trapletti, ``{Multiple anomalous U(1)s
  in heterotic blow-ups},''
  \href{http://dx.doi.org/10.1016/j.physletb.2007.07.007}{{\em Phys. Lett.}
  {\bfseries B652} (2007) 124--127},
\href{http://arxiv.org/abs/0703211}{{\ttfamily arXiv:0703211}}.

\bibitem{Gmeiner:2002es}
F.~Gmeiner, S.~Groot~Nibbelink, H.~P. Nilles, M.~Olechowski, and M.~G.~A.
  Walter, ``{Localized anomalies in heterotic orbifolds},''
  \href{http://dx.doi.org/10.1016/S0550-3213(02)00943-4}{{\em Nucl. Phys.}
  {\bfseries B648} (2003) 35--68},
\href{http://arxiv.org/abs/0208146}{{\ttfamily arXiv:0208146}}.

\bibitem{Green:1984sg}
M.~B. Green and J.~H. Schwarz, ``{Anomaly Cancellation in Supersymmetric D=10
  Gauge Theory and Superstring Theory},''
\href{http://dx.doi.org/10.1016/0370-2693(84)91565-X}{{\em Phys. Lett.}
  {\bfseries B149} (1984) 117--122}.

\bibitem{Schellekens:1986xh}
A.~N. Schellekens and N.~P. Warner, ``{Anomalies, Characters and Strings},''
\href{http://dx.doi.org/10.1016/0550-3213(87)90108-8}{{\em Nucl. Phys.}
  {\bfseries B287} (1987) 317}.

\bibitem{Blaszczyk:2011ig}
M.~Blaszczyk, N.~G. Cabo~Bizet, H.~P. Nilles, and F.~Ruhle, ``{A perfect match
  of MSSM-like orbifold and resolution models via anomalies},''
  \href{http://dx.doi.org/10.1007/JHEP10(2011)117}{{\em JHEP} {\bfseries 1110}
  (2011) 117},
\href{http://arxiv.org/abs/1108.0667}{{\ttfamily arXiv:1108.0667 [hep-th]}}.

\bibitem{Bailin:1999nk}
D.~Bailin and A.~Love, ``{Orbifold compactifications of string theory},''
\href{http://dx.doi.org/10.1016/S0370-1573(98)00126-4}{{\em Phys.Rept.}
  {\bfseries 315} (1999) 285--408}.

\bibitem{Ibanez:1986tp}
L.~E. Ibanez, H.~P. Nilles, and F.~Quevedo, ``{Orbifolds and Wilson Lines},''
\href{http://dx.doi.org/10.1016/0370-2693(87)90066-9}{{\em Phys. Lett.}
  {\bfseries B187} (1987) 25--32}.

\bibitem{Dixon:1985jw}
L.~J. Dixon, J.~A. Harvey, C.~Vafa, and E.~Witten, ``{Strings on Orbifolds},''
\href{http://dx.doi.org/10.1016/0550-3213(85)90593-0}{{\em Nucl. Phys.}
  {\bfseries B261} (1985) 678--686}.

\bibitem{Dixon:1986jc}
L.~J. Dixon, J.~A. Harvey, C.~Vafa, and E.~Witten, ``{Strings on Orbifolds.
  2},''
\href{http://dx.doi.org/10.1016/0550-3213(86)90287-7}{{\em Nucl. Phys.}
  {\bfseries B274} (1986) 285--314}.

\bibitem{Dixon:1986qv}
L.~J. Dixon, D.~Friedan, E.~J. Martinec, and S.~H. Shenker, ``{The Conformal
  Field Theory of Orbifolds},''
\href{http://dx.doi.org/10.1016/0550-3213(87)90676-6}{{\em Nucl. Phys.}
  {\bfseries B282} (1987) 13--73}.

\bibitem{GSW2}
M.~B. Green, J.~H. Schwarz, and E.~Witten, {\em {Superstring Theory}}, vol.~2.
\newblock Cambridge University Press, 1999.

\bibitem{nanathesis}
N.~G. Cabo~Bizet, {\em {Matching the heterotic string in orbifolds and its
  resolutions}}.
\newblock PhD thesis, University of Bonn, November, 2012.

\bibitem{Nilles:2011aj}
H.~P. Nilles, S.~Ramos-Sanchez, P.~K. Vaudrevange, and A.~Wingerter, ``{The
  Orbifolder: A Tool to study the Low Energy Effective Theory of Heterotic
  Orbifolds},'' \href{http://dx.doi.org/10.1016/j.cpc.2012.01.026}{{\em
  Comput.Phys.Commun.} {\bfseries 183} (2012) 1363--1380},
\href{http://arxiv.org/abs/1110.5229}{{\ttfamily arXiv:1110.5229 [hep-th]}}.

\bibitem{Blumenhagen:2005ga}
R.~Blumenhagen, G.~Honecker, and T.~Weigand, ``{Loop-corrected
  compactifications of the heterotic string with line bundles},''
  \href{http://dx.doi.org/10.1088/1126-6708/2005/06/020}{{\em JHEP} {\bfseries
  0506} (2005) 020}, \href{http://arxiv.org/abs/0504232}{{\ttfamily
  arXiv:0504232 [hep-th]}}.

\end{thebibliography}\endgroup
}

\end{document}